# FUSE AND STIS OBSERVATIONS OF INTERVENING O VI ABSORPTION LINE SYSTEMS IN THE SPECTRUM OF PG 0953+415[1,2,3]

B. D. Savage[4], K.R. Sembach[5], T. M Tripp[6], and P. Richter[4]

ABSTRACT

We present Far Ultraviolet Spectroscopic Explorer (FUSE) and Space Telescope Imaging Spectrograph (STIS) observations of the intergalactic medium toward the bright QSO PG 0953+415 ($z_{em}$ = 0.239). The FUSE spectra extend from 905 to 1187 Å have a resolution of 25 km s$^{-1}$ while the STIS spectra cover 1150 to 1730 have a resolution of 7 km s$^{-1}$. Additional STIS observations at 30 km s$^{-1}$ are obtained in selected wavelength ranges. An O VI system at z = 0.06807 is detected in H I Lyα, Lyβ, Lyγ, O VI λλ1031.93, 1037.62, N V λλ1238.82,1242.80, C IV λλ1548.20,1550.77, and C III λ977.02. The observed column densities can be modeled as a low density intervening gas with a metallicity of 0.4 (+0.6, -0.2) times solar in photoionization equilibrium with the ionizing extragalactic background radiation. The best fit is achieved with an ionization parameter, logU = -1.35, which implies $n_H$ ~ 10$^{-5}$ cm$^{-3}$ and a path length of ~ 80 kpc through the absorbing gas. H I Lyα absorption at z = 0.14232 spans a velocity range of 410 km s$^{-1}$ with the strongest components near 0 and 80 km s$^{-1}$ in the z = 0.14232 rest frame. In this system O VI λλ1031.93, 1037.62 absorption is strong near 0 km s$^{-1}$ and not detected at 80 km s$^{-1}$. C III λ977.02 absorption is marginally detected at 80 km s$^{-1}$ but is not detected at 0 km s$^{-1}$. The observations place constraints on the properties of the z = 0.14232 system but do not discriminate between collisional ionization in hot gas versus photoionization in a very low density medium with an ionization parameter logU > -0.74. The z = 0.06807 and 0.14232 O VI systems occur at redshifts where there are peaks in the number density of intervening galaxies along the line of sight determined from WIYN redshift measurements of galaxies in the ~1° field centered on PG 0953+415.

[4] Department of Astronomy, University of Wisconsin-Madison, 475 N. Charter Street, Madison, WI 53706-1582; savage@astro.wisc.edu.

[5] Department of Physics and Astronomy, Johns Hopkins University, Baltimore, MD 21218; sembach@stsci.edu

[6] Department of Astrophysical Sciences, Princeton University, Princeton, NJ, 08544; tripp@astro.princeton.edu.

We combine our observations of PG 0953+415 with those for other QSOs to update the estimate of the low redshift number density of intervening O VI systems. Over a total unobscured redshift path of $\Delta z = 0.43$, we detect six O VI systems with restframe equivalent widths of the O VI λ1031.93 line exceeding 50 mÅ yielding dN/dz = 14 (+9, -6) for $<z>$ = 0.09. This implies a low redshift value of the baryonic contribution to the closure density of the O VI systems of $\Omega_B(O\ VI) > 0.002 h_{75}^{-1}$ assuming the average metallicity in the O VI systems is 0.1 solar.

*Subject headings:* intergalactic medium – quasars: absorption lines – quasars: general – quasars: individual (PG 0953+415)

1. INTRODUCTION

Cosmological simulations predict that a substantial fraction of the baryons at low redshift may exist in shock-heated intergalactic gas with temperatures ranging from $10^5$ to $10^7$ K (Cen & Ostriker 1999a; Davé et al. 1999). While X-ray emission and absorption observations can be used to probe gas with T > $10^6$ K, the ultraviolet (UV) resonance lines of the lithium-like ions O VI, N V, and C IV can be used to search for gas with T < $10^6$ K. The production of these three ions from their lower stages of ionization requires photons or electrons with E > 113.9, 77.5, and 47.9 eV, respectively. Therefore, in the intergalactic medium (IGM) these ions may be created by photoionization by the extragalactic background radiation or by electron collisional ionization in hot gas. In collisional ionization equilibrium, the ions O VI, N V, and C IV peak in abundance at temperatures of (3, 2, and 1)x$10^5$ K, respectively (Sutherland & Dopita 1993). Among these three ions, O VI is the best diagnostic of hot gas in the IGM because of its high production temperature and because of the generally high cosmic abundance of O compared to both C and N.

High spectral resolution observations of the two bright QSOs H 1821+643 ($z_{em}$ = 0.297) and PG 0953+415 ($z_{em}$ = 0.239) with the Space Telescope Imaging Spectrograph (STIS) have revealed that the number density of IGM O VI systems with restframe equivalent widths exceeding 30 mÅ at low redshift is likely very high, with dN/dz > 17 in the general IGM at the 90% confidence level (Tripp, Savage, & Jenkins 2000; Tripp & Savage 2000). A 30 mÅ O VI 1037.62 Å absorption line requires an O VI column density of N(O VI) = 5x$10^{13}$ cm$^{-2}$ if the absorption line is on the linear part of the curve of growth (COG). Assuming an intrinsic oxygen abundance O/H = 0.1 solar and O VI ionization fraction f(O VI) = N(O VI)/N(O) ≤ 0.2 (regardless of the ionization mechanism), the O VI column density implies a column density of ionized hydrogen associated with the O VI system of at least $2.5 \times 10^{18}$ cm$^{-2}$. The number density of O VI systems toward H 1821+643 is so high that an estimate of the lower limit of the baryonic content of the O VI systems assuming an intrinsic oxygen abundance of 0.1 solar yields a cosmological mass density in the units of the current critical density of $\Omega$(O VI) ≥ $0.004 h_{75}^{-1}$. This is comparable to the combined cosmological mass density of stars, cool neutral gas, and X-ray emitting cluster gas at low redshift (Fukugita, Hogan, & Peebles 1998). While this estimate suffers from small number statistics and the uncertain metallicity assumption, it suggests that studies of QSO absorption line systems containing O VI may yield fundamental information about the baryonic content of the low redshift IGM.

In this paper we present FUSE observations of intergalactic absorption toward the bright QSO PG 0953+415 ($z_{em}$ = 0.239) with an emphasis on the O VI absorption line systems. The FUSE observations cover the wavelength range from 905 to 1187 Å with a resolution of ~25 km s$^{-1}$(FWHM). The FUSE observations are combined with STIS E140M echelle observations extending from 1150 to 1730 Å at a resolution of 7 km s$^{-1}$ and G140M grating observations covering 1145 to 1201 Å and 1724 to 1814 Å at a resolution of ~30 km s$^{-1}$. STIS can be used to detect O VI absorption systems with redshifts exceeding ~0.11, while the Far Ultraviolet Spectroscopic Explorer (FUSE) satellite can be used to detect O VI systems from z = 0 to 0.14. Combining STIS and FUSE measurments permits comparisons of the O VI absorption with absorption produced by H I Lyman series lines and the resonance lines of other ions such as C IV, C III, C II, N V, N III, N II, Si IV, Si III, and Si II. The observations of PG 0953+415 extend the previous measurements of the O VI system at z = 0.14232 (Tripp & Savage 2000) and reveal a new O VI system at z = 0.06807 that is also detected in H I Lyα, Lyβ, Lyγ, N V λλ1238.82, 1242.80, C IV λλ1548.20, 1550.77, and C III λ977.02.

The FUSE and STIS observations and data handling procedures are discussed in §2. The IGM absorption line results are given in §3. The physical conditions in the two intervening O VI systems clearly detected toward PG 0953+415 are discussed in §4. The possible origins of the O VI systems and their relationship to other structures along the line of sight are discussed in §5. We update the estimate of the number density of O VI systems at low redshift and the implications for the baryonic content of these systems in §6. The results are summarized in §7.

2. OBSERVATIONS

2.1 FUSE Observations

FUSE contains two microchannel plate detectors and four-coaligned optical channels. Two channels are optimized for short wavelengths (SiC, 905 -1100Å) and two are optimized for longer wavelengths (Al+LiF, 1000 -1187Å). Descriptions of the instrument and its on-orbit performance are given by Moos et al. (2000) and Sahnow et al. (2000). PG 0953+415 was centered in the large (30"x30") aperture of the LiF1 channel of detector 1 for 15 successful exposures totaling 34 ksec on 30 December 1999 and for 19 successful exposures totaling 38 ksec on 4 May 2000. The LiF1 channel was used for guiding on the QSO. During the December 1999 integrations the SiC1 and LiF2 channels were reasonably well aligned, but the total exposure time obtained in the SiC2 channel was only 21.8 ksec because of mis-alignment problems. During the May 2000 observations, the alignment was much better with all channels yielding total integration times ranging from 34.6 to 38 ksec. The individual exposures are stored in the FUSE archive at the Multi-Mission Archive at the Space Telescope Science Institute under the identifications P1012201 (exposures 1-3 and 6-17) and P1012202 (exposures 3-21).

The processing made use of the FUSE calibration pipeline available at the Johns Hopkins University as of June 2000 (CALFUSE Version 1.7.5). The processing steps applied to the time-tagged photon event lists are identical to those discussed by Sembach et al. (2001a). The processing screens the observations for Earth limb avoidance and passage through the South Atlantic Anomaly. Detector event bursts were identified and removed as a separate processing step. Corrections are applied for geometric distortions, spectral motions caused by thermal changes, and Doppler shifts caused by orbital motions. Suitable astigmatism and flatfield corrections were not available when the observations were processed. The December 1999 and May 2000 spectra were weighted according to the signal-to-noise and averaged after shifting for velocity offsets. The presence of fixed-pattern noise was evaluated by comparing the spectra produced by the different channels. We required absorption lines be present in the data for two channels to be considered positive detections.

Scattered light and detector backgrounds are negligible at the flux levels recorded for PG 0953+415. However, some wavelengths are significantly affected by airglow emission lines from the Earth's atmosphere which are stronger during the day. Separate extractions were performed on the night-only observations versus the day + night observations to evaluate the impact of these airglow emissions, particularly in the lines of O I and N I.

The final processed (day+night) spectra for the two most sensitive channels in each wavelength region are shown in Figure 1. Below 1000 Å we illustrate spectra for the more sensitive SiC channels. Longward of 1000 Å we illustrate spectra for the LiF channels except for the short wavelength region from 1075 to 1091 Å where SiC spectra are displayed. The signal-to-noise and resolution vary as a function of wavelength and from channel to channel. Because of differences in resolution and the existence of fixed-pattern noise, we have found it helpful to analyze the independent spectra from different channels rather than to simply combine the spectra. It is often true based on signal-to-noise, resolution, and fixed pattern noise considerations that the measurements from a single channel are superior to those from the other channels or for all the channels combined. The signal-to-noise values per resolution element actually achieved in the combined spectra are: 6 and 9 at 950 Å in the SiC1B and SiC2A channels; 16 and 12 at 1050 Å in the LiF1A and LiF2B channels; and 12 and 15 at 1150 Å in the LiF1B and LiF2A channels.

The resolution achieved in these observations ranges from ~ 17 to 25 km s$^{-1}$ (FWHM) depending on the wavelength and the channel. There was no difference in resolution from the December 1999 to the May 2000 observations. The wavelengths have a 1$\sigma$ accuracy equivalent to a velocity accuracy of ~ 6 km s$^{-1}$ averaged over an entire channel. However, there are 5 -10 Å regions of the spectrum where errors as large as 10 to 15 km s$^{-1}$ may be present. The zero point of the wavelength scale was set by comparing ISM lines observed by FUSE with those recorded by much higher resolution STIS observations from Tripp & Savage (2000).

2.2 STIS E140M Observations

The full details of the STIS observations of PG 0953+415 obtained with the E140M echelle mode are reported by Tripp & Savage (2000) and Tripp et al. (2002). Information on the design and performance of STIS are found in Woodgate et al. (1998) and Kimble et al. (1998). The spectra cover the wavelength range from 1150 to 1730 Å with an average S/N of ~7 at a resolution of ~7 km s$^{-1}$ (FWHM). Tripp & Savage (2000) and Tripp et al. (2002) present results for the IGM absorption lines seen in the STIS spectra, while Fabian et al. (2002) discuss the ISM absorption features. Our specific interest here is in the IGM absorption features presented by Tripp et al. (2002). That paper lists the observed extragalactic absorption lines, their identifications, wavelengths, rest frame equivalent widths, and the integrated apparent column densities using the software of Sembach & Savage (1992). They also give the Doppler parameters and column densities derived from Voigt profile fitting to the STIS absorption line data using the using the software of Fitzpatrick & Spitzer (1997) and the line spread functions for the E140M mode with the 0.2"x0.2" aperture given in the STIS Cycle 10 Handbook (Leitherer et al. 2000).

The measured wavelengths, redshifts, IDs, restframe wavelengths, restframe equivalent widths, and apparent column densities or upper limits based on these previous STIS E140M observations are given in Table 1 for the absorbing systems at z = 0.06807 and 0.14232.

2.3 STIS G140M and G230M Observations

The STIS E140M observations of PG 0953+415 have been supplemented with G140M observations of the wavelength region from 1145 to 1201 Å and with G230M observations of the region from 1724 to 1814 Å. These 13.5 ksec first order grating integrations with the 0.2"x 50" slit have a resolution of ~ 30 km s$^{-1}$. The G140M observations were designed to provide higher S/N observations of the important Ly$\beta$ and O VI lines in the z = 0.14232 system. The G230M observation was designed to search for C IV $\lambda\lambda$1548.20, 1550.77 absorption in the O VI system at z = 0.14232.

The STIS G140M and G230M observations were processed using the STIS science team software (Lindler 1998). Standard procedures were used for flatfielding and wavelength calibration. Two 50-pixel wide regions centered 52 pixels away from the spectrum were used to determine the background. The total integration time of 13.5 ksec was broken up into five exposures. These were separately extracted and combined by weighting them inversely by their variances averaged over a large portion of the spectrum.

3. ABSORPTION LINE RESULTS

3.1 FUSE

FUSE spectra of PG 0953+415 are shown in Figure 1. Definite ISM and IGM absorption lines are marked with the numbered tic marks. Their identifications are listed to the right of each panel. For a line to be considered definite we require a measured equivalent width of >3$\sigma$ significance in the spectra of two or more channels. This results in an overall significance level of > 4.3$\sigma$. The possible presence of fixed-pattern noise in the spectra lead us to adopt these conservative criteria for the identification of absorption lines. At some of the wavelengths where we have searched for O VI features and have not found a definite absorption feature, we have placed a numbered tick mark and but add a question mark next to the wavelength listing the feature of interest in the line lists shown in Figure 1.

The absorption line equivalent width measurements and their errors were determined using the methods described by Sembach & Savage (1992). The errors include continuum placement uncertainties and the statistical photon counting errors.

Many of the absorption lines identified in Figure 1 are produced in the Galactic interstellar medium (ISM) and are not discussed further in this paper unless they blend with IGM absorption lines. Measurements for the IGM absorption lines detected in the FUSE spectra are listed in Table 2, which gives the average observed wavelength, the redshift, the identification, the restframe wavelength of the identified species in Å, the restframe equivalent width and ±1$\sigma$ error for the absorption in mÅ. Equivalent widths are listed for lines detected in the two LiF channels. The notes to Table 2 provide information about various difficulties associated with the measurements or identifications.

The IGM lines detected in the FUSE observations include 2 Ly$\beta$ lines that confirm the STIS Ly$\alpha$ detections reported by Tripp et al. (2002). In addition Ly$\gamma$ is detected in the H I system at z = 0.06807. In that system we also detect C III $\lambda$977.02 and O VI $\lambda\lambda$1031.93, 1037.62. The FUSE observations confirm the presence of the O VI absorber at z = 0.14232 discussed by Tripp & Savage (2000). In that same system the FUSE observations also reveal the presence of Ly$\beta$ blended with Ly$\delta$ from a strong system at z = 0.23349 associated with the QSO. C III $\lambda$977.02 is marginally detected (2.6$\sigma$) at z = 0.14265, corresponding to the redshift of the second strongest Ly$\alpha$ line in this multi-component H I system (see § 4.2).

### 3.2. STIS G140M and G230M Results for the z = 0.14232 system

We list at the bottom of Table 1 the G140M and G230M measurements (or upper limits) for the lines of Ly$\beta$, O VI $\lambda\lambda$1031.93, 1037.62, and C IV $\lambda\lambda$1548.20, 1550.77 in the multi-component system at z = 0.14232. The G140M and G230M measurements confirm the detection of Ly$\beta$ in the system at z = 0.14232 blended with Ly$\delta$ in the system at z= 0.23349. O VI $\lambda\lambda$1031.93, 1037.62 in the z = 0.14232 system has maximum absorption near v = 0 km s$^{-1}$ with weaker absorption possibly extending to 100 km s$^{-1}$. C IV $\lambda$1548.20 is not convincingly detected at z = 0.14232 or at z = 0.14262. However, a feature of 2$\sigma$ significance does appear near z = 0.14262 (see Table 1, note 8). The complex velocity structure and physical conditions in this multicomponent system are discussed in §4.2.

### 3.3 A Search for Ly$\beta$, O VI, and C III at the redshifts of the Ly$\alpha$ Lines

Understanding the origin of O VI systems will require information about which Ly$\alpha$ systems have associated O VI absorption and which do not. Table 3 lists the redshifts and restframe equivalent widths of the Ly$\alpha$ lines detected by STIS (Tripp et al. 2002). For each of the Ly$\alpha$ lines listed we have searched the FUSE and STIS observations for Ly$\beta$, O VI $\lambda\lambda$1031.93, 1037.62, and C III $\lambda$977.02. The definite detections are indicated in column 3 of Table 3. The notes to the table indicate the numerous cases where the lines we are searching for blend with an ISM or unrelated IGM absorption line. When the blending affects one or both of the O VI doublet lines we have added the note " blend O VI" to column 3 of Table 3. The large number of blends is caused primarily by overlap with Milky Way H$_2$ lines. While strong intervening O VI absorption might be noticeable in some of these blends, weak O VI absorption may be hidden. When no detection is indicated, the 4$\sigma$ limiting restframe equivalent widths in the absence of blending are typically ~60 mÅ for the FUSE observations and ~40mÅ for the STIS observations.

Ignoring the associated systems at z = 0.22527, 0.23260, and 0.23349, we see from Table 3 that Ly β is detected in three Lyα systems, the O VI doublet is found in two systems, and C III is found in two systems. Blending confuses the search for one or both of the O VI lines in eight of the twenty Lyα line systems. In this accounting the multicomponent Lyα system extending from z = 0.14232 to 0.14335 has been counted as a single system.

## 4. PHYSICAL CONDITIONS IN THE O VI SYSTEMS

In this section we discuss the physical conditions and metallicity constraints determined for the two O VI systems detected by FUSE and STIS in the spectrum of PG 0953+415.

### 4.1 The O VI system at z = 0.06807

The FUSE and STIS line profiles for the O VI system near z = 0.06807 are displayed on a velocity basis in Figures 2 and 3. This O VI system is detected in the lines of Lyα, Lyβ, Lyγ, O VI λλ1031.93, 1037.62, N V λλ1238.82,1242.80, C IV λλ1548.20, 1550.77, and C III λ977.02. In addition, upper limits exist for Si IV λ1393.76, Si III λ1206.50, Si II λ1260.42, and C II λ1335.53. These measurements permit an evaluation of the ionization conditions and physical conditions in the absorbing gas. The system is quite highly ionized given the strength of the N V and O VI lines relative to the lower ionization lines.

The column densities or limits we have obtained for these ions are listed in Table 4. In the case of H I, the Voigt profile fit to the Lyα line shown in Figure 3 yields b = 23± 3 km s$^{-1}$ and logN(H I) = 14.35 (+0.08, -0.14). An alternate approach for estimating logN(H I) is to fit the measured rest frame Lyα, Lyβ, and Lyγ equivalent widths to a single component COG. The result is b = 22.5 (+2.8, -2.5) km s$^{-1}$ and logN(H I) = 14.39 (+0.11, -0.10). For C IV and N V we have adopted the Voigt profile fit results from Tripp et al. (2002). These fits are shown on the observed profiles in Figure 3. The Doppler parameters obtained from these fits are b = 8 (+7, -4) km s$^{-1}$ for C IV and b = 10±4 km s$^{-1}$ for N V. The column density of C III was obtained by fitting the measured rest-frame equivalent width to a single component COG with the value of b = 8 (+7, -4) km s$^{-1}$ obtained from the C IV profile fit. The 4σ column density upper limits for C II, Si II, Si III, and Si IV assume the absorption occurs in a component with b = 8 km s$^{-1}$. For O VI, a fit of the measured equivalent widths to a single component COG yields logN(O VI) = 14.22 (+0.24, -0.15) and b = 17 (+13, -6) km s$^{-1}$.

For pure thermal line broadening the standard Gaussian line spread parameter, b, is related to gas temperature through b (km s$^{-1}$) = (2kT/m)$^{1/2}$ = 0.129 (T/A)$^{1/2}$, where A is the atomic weight. Therefore, the values of b obtained for H I (from profile and COG fitting), C IV (from profile fitting), N V (from profile fitting), and O VI (from the COG) imply $1\sigma$ upper limits to the temperature of the gas of T (H I) < $4.1 \times 10^4$ K , T(C IV) < $1.6 \times 10^5$ K , T(N V) < $1.6 \times 10^5$ K , and T(O VI) < $8.7 \times 10^5$ K . Allowing for the effects of turbulence, multiple components, and gas flows would lower all these temperature limits. The H I temperature limit implies the H I exists in warm gas. The more highly ionized species could exist in the same warm gas if they are produced by photoionization by the extragalactic background. However, the temperature limits are also consistent with the possibility that the highly ionized species are created by collisional ionization in equilibrium or non-equilibrium processes in a hotter gas phase.

If H I, C IV, and N V are produced in the same warm gas, then the decrease in the observed value of b from the light to the heavier ions can be used to separately determine the temperature of the gas and the amount of turbulent broadening. If the line of sight turbulent broadening is described by a Gaussian function which can be added in quadrature to the thermal Doppler broadening, then $b_{obs}^2 = b_{thermal}^2 + b_{turb}^2 = 0.01664 \, (T/A) + b_{turb}^2$. Among the high ions, the most accurate estimate for b is for N V. Taking b(H I) = 23 km s$^{-1}$ and b(N V) = 10 km s$^{-1}$, we find the observed H I and N V profile widths to be consistent with gas at T ~ 28,000 K with a turbulent broadening of $b_{turb}$ ~ 8.2 km s$^{-1}$.

We explore the possibility that the species detected by FUSE and STIS in the z = 0.06807 absorber are created by photoionization. We employ the equilibrium photoionization code CLOUDY (version 90.04; Ferland et al. 1998) and treat the absorber as a constant density plane-parallel slab photoionized by the integrated background radiation from QSOs and AGNs with a relative energy distribution as determined by Haardt & Madau (1996) for z = 0.06. We set the mean intensity at the Lyman limit $J_\nu$(LL) = $1 \times 10^{-23}$ ergs cm$^{-2}$ s$^{-1}$ Hz$^{-1}$ sr$^{-1}$, which is in agreement with observational constraints (Kulkarni & Fall 1993; Maloney 1993; Tumlinson et al. 1999; Davé & Tripp 2001) and theoretical expectations (Shull et al. 1999). We then varied the ionization parameter, U = $n_\gamma / n_H$ = ionizing photon density / total H number density, and the gas metallicity to find a model that best matches the observed column densities and limits given in Table 4. For metallicities we express the linear abundance of element X to Y as (X/Y) and the logarithmic abundance relative to solar in the standard notation, [X/Y] = log (X/Y) – log (X/Y)$_O$. The relative heavy element abundances were assumed to be the solar values from Grevesse & Anders (1989) and Grevesse & Noels (1993) but we varied the overall metallicity as indicated by (M/H) or [M/H].

The results of the photoionization modeling are shown in Figure 4 for the O VI system at z = 0.06807. The various curves show the high ion column densities predicted by the models as a function of logU for logN(H I) = 14.39 and [M/H] = -0.4. The measurements are plotted at logU = -1.35 where the best fit is achieved for the column densities of O VI, N V, and C IV. Acceptable fits (consistent with the ±1σ error bars) are achieved over the ranges logU = -1.1 and [M/H] = -0.7 to log U = -1.6 and [M/H] = 0.0.

If these species are created by photoionization by the extragalactic background, then the metallicity is large with a best fit result of 0.4 (+0.6, -0.2) times solar. Nitrogen is often found to be underabundant in low metallicity gas, a result usually interpreted as being due to the secondary nucleosynthetic origin of nitrogen. The strength of the N V absorption relative to the O VI and C IV absorption is consistent with an origin of these ions in relatively high metallicity gas.

The total (neutral+ ionized) hydrogen density, $n_H$ (cm$^{-3}$) in the gas is indicated at the top of Figure 4. For the best fit value of $\log U = -1.35$, $n_H \sim 10^{-5}$ cm$^{-3}$, and f(O VI), the fraction of O in the form of O VI, is 0.15. To produce the observed O VI column density in a gas with 0.4 times solar abundances requires an absorption path length of ~80 kpc.

In §5 we show that the z = 0.06807 absorber is associated with a peak in the redshift distribution of galaxies in the ~1° field centered on PG 0953+415. The absorber therefore is probably associated with a group of galaxies or a large scale structure. The observed abundance of 0.4 (+0.6, -0.2) times solar is similar to the uniform abundance of 0.32 times solar found for X-ray emitting intragroup gas at low redshift (Fukazawa et al. 1998). Looking locally, the absorber has an abundance similar to that found in the Magellanic Stream (0.2 to 0.4 times solar, Lu et al. 1998; Sembach et al. 2001b) but an abundance approximately four times larger than that found in the high-velocity cloud Complex C (~0.1 times solar, Wakker et al. 1999; Richter et al. 2001b).

We cannot rule out the possibility that the O VI, N V, and C IV in the z = 0.06807 absorber are produced in hot gas with T ~ (1-3) x10$^5$ K. The +1$\sigma$ upper limits to the high ionization line widths are consistent with such an explanation. However, the 1$\sigma$ upper limit to the H I line width requires that the hydrogen absorption arise in a warm medium with T < 4.1x10$^4$ K. Therefore, a hot gas explanation requires a multiphase absorber with H I tracing a warm phase and the highly ionized species tracing a hotter phase.

It is interesting that the observed column densities of O VI, N V, and C IV in the z= 0.06807 system are almost identical to the column densities of these three species observed for absorption paths through the Milky Way halo. Savage et al. (1997, 2000) estimate logN(O VI) = 14.22±0.06, logN(N V) = 13.43±0.09, and logN(C IV) = 14.08±0.07 for the high ionization absorption line column densities perpendicular to the Galactic plane through the entire halo to one side of the Galactic disk. These Milky Way absorption features are believed to arise in the cooling gas of a Galactic fountain. However, through the Milky Way halo, the value of logN(Si IV) = 13.56±0.06, which is much larger than the 4$\sigma$ upper limit of <12.6 found for the z = 0.06807 system. Also, for hot halo gas in the Milky Way, the effective b values found for C IV are typically several times larger than in the z = 0.06807 system (Sembach & Savage 1992; Savage et al. 1997). These differences suggest the ionization mechanisms controlling the relative ionic abundances in the z = 0.06807 system are not the same as those operating in the Milky Way halo.

### 4.2 The O VI System at z = 0.14232

The STIS observations of the z = 0.14232 absorber by Tripp & Savage (2000) reveal a complex multi-component H I Lyα profile containing 6 components spread over ~410 km s$^{-1}$. Both lines of O VI were detected but N V and the lines of Si III and C II were not detected. The O VI lines were recorded in a noisy region of the STIS spectrum. Tripp & Savage remarked that the O VI λ1037.62 line was broader than the larger f-value O VI λ1031.9 line. Although they reported results for both lines, they prefered the results for the stronger λ1031.93 Å line and suggested that the λ1037.62 line might be affected by detector fixed-pattern noise. Our new FUSE and STIS observations allow us to better understand the complex multi-component nature of this absorption system.

In Figure 5 we show the continuum normalized FUSE absorption profiles for the z = 0.14232 absorption system on a restframe velocity basis. From top to bottom we show the LiF1 and LiF2 profiles for O VI λλ1031.93, 1037.62, C III λ977.02, Lyγ (no detection), and Lyβ. In Figure 6 we show the STIS E140M observation of Lyα along with the new STIS G140M observations of Lyβ, O VI λλ1031.93, 1037.62, and C IV λλ1548.20, 1550.77 (no detection).

There are various problems associated with understanding this inhomogeneous set of spectra. The Lyβ absorption corresponding to the z = 0.14232 system is contaminated by Lyδ absorption at z = 0.23349 with logN(HI) = 14.60±0.02 and b = 23±1 km s$^{-1}$ (Tripp et al. 2002). The contaminating line is expected at 1171.50 Å, which in the z = 0.14232 reference frame for Lyβ occurs at a velocity of -51 km s$^{-1}$. Using the value of logN(H I) and b for the associated z = 0.23349 system determined from Lyα and Lyβ, the Lyδ line should have an observed equivalent width of 45 mÅ which can be compared to the value $W_{obs}$ = 78±13 mÅ observed for the Lyδ (z=0.23350) and Lyβ (z = 0.14232) blend with the G140M measurement or the error weighted average $W_{obs}$ = 78±18 obtained from the FUSE LiF1 and LiF2 observations. This contamination explains why the Lyβ profiles for the z = 0.14232 system shown in Figures 4 and 5 extend to more negative velocity than Lyα.

The FUSE and STIS (old and new) observations of O VI are consistent with maximum O VI absorption near v = 0 km s$^{-1}$. Weak O VI absorption possibly extends from 40 to 100 km s$^{-1}$ but the different observations are inconsistent over this velocity range. O VI absorption is not detected at 80 km s$^{-1}$, the velocity of the second strongest Lyα component. However, absorption by C III is marginally seen at that velocity. The possible 2.6σ detection of C III associated with this second Lyα component implies a multi-phase nature for the collection of H I absorbing components in this system. The absorber at 0 km s$^{-1}$ containing O VI is highly ionized given the presence of H I, O VI, and the limits on N V, C IV, C III, and other species. The Lyα absorber at 80 km s$^{-1}$ possibly contains C III, but no clear evidence for O VI or C IV. The column densities in the various components of the multicomponent system at z = 0.14232 are listed in Table 5.

With better information for O VI and C IV, our new observations further constrain the origin of the ionization in the component at 0 km s$^{-1}$ compared to the constraints discussed by Tripp & Savage (2000). The new value of the O VI column density and the limits on N V and C IV are consistent with photoionization in a very low density medium with logU > -0.74 and $n_H$ < 10$^{-5.7}$ cm$^{-3}$. The photoionization model curves and data points are shown in Figure 7. The model calculation is the same as for Figure 4 (see § 4.1) except for adopting the Haardt & Madau (1996) background radiation field for z = 0.12 which is close to the redshift of the system. We used the estimated radiation field for z = 0.12 because a digital tabulation of the field for z = 0.14 was not available. To match the observed value of N(O VI) and N(H I) with logU = -0.74 and $J_\nu$(LL) = 10$^{-23}$ ergs$^{-1}$ s$^{-1}$ cm$^{-2}$ Hz$^{-1}$, the photoionization model requires a metallicity ~ 0.4 times solar and a path length of ~420 kpc. If the value of logU is as large as -0.1, the required metallicity decreases to ~0.2 times solar and the density decreases to $n_H$ ~ 10$^{-6.3}$ cm$^{-3}$.

The absorption is also consistent with O VI arising from equilibrium or non-equilibrium collisional ionization in a hot medium. For gas in collisional ionization equilibrium the value of N(O VI) and the limits on N(N V) and N(C IV) require T > 2.5x10$^5$ K. However, in this case the H I and O VI absorption would need to occur in a multiphase medium since the moderate b value for the H I absorption implies T < 58,000 K.

The amount of information about the absorber at z = 0.14265 is rather limited with logN(H I) = 13.43±0.04, logN(C III) = 12.6±0.2, and 4σ limits for logN(C IV) and logN(O VI) of 13.3 and 13.8, respectively. The b value for the H I absorption of 30±9 km s$^{-1}$ yields a limit to the temperature of the absorber of < 5.4x10$^4$ K with a 1σ limit of < 9.1x10$^4$ K. To study the possible photoionization of this gas we adopt the model we used to study the absorber at z = 0.14232 but have adjusted for the lower H I column density of logN(H I) = 13.43±0.04. For the same abundance required to fit the data for the z = 0.14232 absorber, [M/H] = -0.4, and the values of logN(O VI) < 13.8 and logN(C III) = 12.6 - 1σ = 12.4, the measurements only roughly constrain the ionization parameter to lie somewhere between logU = -1.0 and -2.4 with best fit values of logU = -1.25 and -2.15 for logN(C III) = 12.6. If the two absorbers are photoionized, the one at z = 0.14265 is occurring in somewhat denser gas than the one at z = 0.14232 which is constrained to logU > - 0.74.

In §5 we show that the absorbers near 0.14232 occur close to the redshift where there is a peak in the number density of galaxies in the one degree field centered on PG 0953+415. Therefore, the number of possible explanations for the absorbers is large since the six H I absorbers spanning the 410 km s$^{-1}$ velocity range may be recording very different absorbing structures in a galaxy group including for example cooling group gas, galaxy halos, and tidally stripped gas. It is also possible the absorption arises in a large scale network of shock heated gas filaments that is associated with the galaxies along the line of sight.

## 5. RELATIONSHIPS BETWEEN O VI ABSORPTION SYSTEMS AND OTHER STRUCTURES TOWARD PG 0953+415

We have used the WIYN observatory fiber-fed multi-object spectrograph to measure redshifts for galaxies with B magnitudes brighter than approximately 19 in the ~1$^{o}$ field centered on PG 0953+415. The methods are identical to those discussed by Tripp et al. (1998).

Figure 8 shows the histogram of galaxy redshifts in the one degree field. The redshifts of Lyα absorbers are indicated with vertical lines whose lengths scale with the absorber rest equivalent widths. The O VI absorbers at z= 0.06807 and z = 0.14232 are also marked in the Figure. The O VI systems lie near strong enhancements in the density distribution of galaxies as a function of redshift.

Although there is an excess of galaxies near the redshift of the O VI absorber at z = 0.06807 (see Figure 8), there is no galaxy in the WIYN study particularly close to the actual line of sight to the QSO. The five galaxies representing the peak in the histogram range in z from 0.06792 to 0.06882 ( Δv = -42 to +210 km s$^{-1}$) and have projected distances ranging from 0.92 to 3.5 Mpc for H$_o$ = 75 km s$^{-1}$ and q$_o$ = 0.0. Information about the five galaxies is summarized in Table 6. These are all luminous galaxies (0.6 –4.0L$_*$).

The multi-component Lyα absorber near z = 0.14232, which has strong O VI absorption at the redshift of the strongest Lyα component, is also associated with a maximum in the redshift distribution of galaxies in the one degree field centered on the QSO. The details are provided in Table 1 and §2.2 of Tripp & Savage (2000). A galaxy is found at a projected distance of 0.395 Mpc separated by ~130 km s$^{-1}$ from the O VI absorber, and three additional galaxies are found within 130 km s$^{-1}$ with projected separations ranging from 1.0 to 3.0 Mpc. All four galaxies are luminous (0.6 –4.0L$_*$) and two show [O II] emission implying active star formation.

The two definite O VI absorbers toward PG 0953+415 occur at redshifts that coincide with two of the four most pronounced peaks in Figure 8. Although a Lyα line at z = 0.0451 with W$_r$ = 37±6 mÅ is found near the pronounced grouping of galaxies with redshifts centered at approximately 0.041, there is no O VI absorption near the redshift of that Lyα absorber. Therefore, the most pronounced grouping of galaxies along the line of sight has no related O VI absorber.

The observed association of the O VI absorbers toward PG 0953+415 with peaks in the galaxy distribution but not with individual galaxies close to the line of sight is a result affected by observational bias since the WIYN redshift survey did not extend to faint galaxies in the general direction of PG 0953+415. Also, several quite bright stars in the field of PG 0953+415 make it difficult to identify galaxies in their vicinity. Redshift surveys extending to fainter galaxies in the field of PG 0953+415 would be valuable to establish if this result extends to lower luminosity galaxies. Nevertheless, it is interesting to note that this association of O VI absorbing gas with galaxy structures appears to be qualitatively consistent with the most recent hydrodynamical simulations of structure formation (Cen et al. 2001). In those simulations, the O VI bearing gas is found to be in a somewhat connected network of filaments that tracks the galaxy distribution (see their Figure 1; see also Figure 4 in Cen & Ostriker 1999b), but on the basis of the absorber overdensities, the O VI systems are predicted to be located at relatively large distances from virialized groups and clusters. Deeper redshift surveys of the field of PG0953+415 (and other QSO fields) would therefore also provide a valuable test of the cosmological simulations.

Turning to the sight line to the QSO H 1821+643 studied by Tripp et al. (2000) and Oegerle et al. (2000), we find that the O VI absorbers at $z = 0.12137$, $0.22497$, $0.24531$, and $0.26659$ with $W_r$ (1031.93) exceeding 50 mÅ lie at redshifts where galaxies have been detected along the line of sight with impact parameters of less than 0.8 Mpc (see Table 7 in Tripp et al. 1998). The two strongest O VI systems at $z = 0.12137$ and $z = 0.22497$ lie near the redshifts of the two galaxies with the smallest impact parameters so far found for the line of sight. The $z = 0.12137$ system lies at a projected distance of 0.144 Mpc from a galaxy at $z = 0.12154$, while the $z = 0.22497$ system lies 0.105 Mpc from a galaxy at $z = 0.22560$ (see Table 7 in Tripp et al. 1998). This suggests a possible origin of the two strongest O VI systems in the distant outskirts of the halos of galaxies along this line of sight. Gas at 0.1 Mpc distances from galaxies could be the result of tidal interactions or gas left over from the galaxy formation process. Local examples of such gas near the Milky Way are the Magellanic Stream and Complex C. It is interesting that both of these local structures have strong associated O VI absorption (Sembach et al. 2000).

The strong O VI absorbers toward PG 0953+415 and H 1821+643 occur at redshifts that imply an association with galaxy groups or large scale structures. Thus, the cross-section subtended by galaxy groups in O VI absorption may be relatively large. However, to make firm statements it will be necessary to build up the statistical base by obtaining similar observations for other sight lines. The association of the strong O VI absorbers toward PG 0953+415 and H 1821+643 with galaxy groups along the line of sight provides strong evidence that these particular absorbers are intervening systems and are not systems that have been ejected from the quasars.

## 6. THE NUMBER DENSITY OF O VI SYSTEMS AT LOW REDSHIFT

Although the overall redshift path is still quite small, the observations for PG 0953+415, H 1821+643 (Tripp et al. 2000), PG 0804+761 (Richter et al. 2001a), and 3C 273 (Sembach et al. 2001b) allow us to update the estimate of the number density of O VI systems at low redshift and the implications of that estimate for the low redshift baryonic content of the O VI systems.

In Table 7 we list the number of intervening Ly$\alpha$ and O VI absorption systems detected at low redshift in STIS and FUSE data with rest frame equivalent widths for the O VI $\lambda$1031.93 line exceeding 50 mÅ. Absorbers within 5000 km s$^{-1}$ of $z_{em}$ are excluded to avoid contamination from associated absorbers. In the case of H 1821+643, PG 0953+415, and 3C 273 the results are based on careful searches for O VI systems at the redshifts of known Ly$\alpha$ systems. The table does not include two weak O VI systems toward H 1821+643 and one weak system toward 3C 273 with $W_r < 50$ mÅ. For PG 0804+761 existing UV data with the HST cover only a small redshift path. Therefore, more than one Ly$\alpha$ line may exist. For PKS 0405-1219, Chen & Prochaska (2001) have reported on the analysis of one strong O VI system but did not comment on other possible O VI systems  The total absorption path corrected approximately for spectral regions blocked by ISM lines or absorption lines from other IGM systems is given in column 3 of Table 7. Footnote a of Table 7 describes the spectral blocking estimate.

Combining the results for H 1821+643, PG 0953+415, PG 0804+761, and 3C 273 we find that six O VI systems with $W_r > 50$ mÅ have been detected over an unblocked redshift path of 0.43, which implies dN/dz = 14 (+9, -6) for low redshift O VI systems. The $\pm 1\sigma$ error estimate allows for small sample statistics according to Gehrels (1986). This value of dN/dz for O VI systems with $W_r > 50$ mÅ at $\langle z \rangle = 0.09$ can be compared to the low z estimate dN/dz > 17 (90% confidence level) derived by Tripp et al. (2000) for O VI systems with a smaller equivalent width limit, $W_r > 30$ mÅ, and a larger average redshift, $\langle z \rangle \sim 0.2$.

The mean cosmological mass density in the O VI absorbers, in units of the current critical density, $\rho_c$, can be estimated using

$$\Omega_b(\text{O VI}) = (\mu m_H H_o/\rho_c c)\ f(\text{O VI})^{-1}\ (H/O)\ \Sigma N(\text{O VI})_i / \Sigma \Delta X_i, \quad (1)$$

where $\mu$ is the mean molecular weight (taken to be 1.3), f(O VI) is the O VI ionization fraction, (H/O) is the assumed hydrogen to oxygen ratio in the O VI systems by number, $\Sigma N(\text{O VI})_i$ is the total O VI column density summed over the absorbers found in the absorption distance interval $\Sigma \Delta X_i$ (Bahcall & Peebles 1969; Burles & Tytler 1996), corrected for blocked spectral regions. For $q_o = 0$ and a redshift range from $z_{min}$ to $z_{max}$, $\Delta X = 1/2\{\ [(1+z_{max})^2 -1] - [(1+z_{min})^2 -1]\}$. In our case $z_{min} = 0$ and therefore, $\Delta X = z_{max} + z_{max}^2 / 2$. To set a conservative lower limit on $\Omega_b$(O VI), we assume [O/H] = -0.4, and f(O VI) = 0.2 which is close to the maximum value found in photo- or collisional-ionization equilibrium (Tripp & Savage 2000). Using the values of $\Sigma N$(O VI) from Table 6 and accounting approximately for spectral blockage by ISM lines and IGM lines from other absorption systems (see footnote a of Table 7), we obtain $\Omega_b$(O VI) $\geq$ 0.0004$h_{75}^{-1}$. If the oxygen abundance in the O VI systems is instead [O/H] = -1.0, we obtain $\Omega_b$(O VI) $\geq$ 0.002$h_{75}^{-1}$ for $<z>$ = 0.09. The lower limit obtained here is several times smaller than the estimate of Tripp et al. (2000) based on the STIS data for H1821+643 alone. For comparison the cosmological mass density of stars in galaxies at low redshift is reported by Fukugita et al. (1998) to be $\sim$ 0.0033$h_{75}^{-1}$ at z = 0.0.

To improve upon these estimates it will be necessary to reduce the statistical errors by finding more systems, obtaining better information about the numerous systems with $W_r < 50$ mÅ not included in this estimate, and more accurately evaluating the effects of spectral line blocking. In addition, obtaining a better understanding of the ionization mechanisms and determining the oxygen abundances in representative O VI systems will be crucial for eventually evaluating the overall baryonic content of these systems. However, even in the absence of such work it is evident that the O VI systems represent an important component of the baryonic matter in the low redshift universe. Their study should therefore provide significant clues about the efficiency of the conversion of gaseous matter in the evolving universe into stars and galaxies.

7. SUMMARY

Spectra obtained of the bright QSO PG 0953+415 (z = 0.239) with FUSE and STIS are analyzed in order to study intervening intergalactic O VI absorption systems. The observations provide nearly complete wavelength coverage from 915 to 1800 Å with resolution ($\lambda/\Delta\lambda$) of $\sim$ 15,000 from 915 to 1187 Å and $\sim$45,000 from 1150 to 1730 Å. Two additional STIS integrations at 10,000 resolution are included in the study.

1. A search for O VI absorption at the redshifts of 20 intervening Ly$\alpha$ absorbing systems yields two definite O VI systems, 10 non-detections, and 8 cases where one or both of the O VI $\lambda\lambda$1032.93, 1027.62 lines blends with ISM or unrelated IGM absorbers.

2. The observations reveal a strong absorption system at z = 0.06807 which is detected in Ly$\alpha$, Ly$\beta$, Ly$\gamma$, O VI $\lambda\lambda$1031.93, 1037.62, N V $\lambda\lambda$1238.80, 1242.80, C IV $\lambda\lambda$1548.20, 1550.77, and C III $\lambda$977.02. The measurements yield column densities for all the detected species and upper limits for other ions.

3. The observed column densities and line widths for H I, O VI, N V, C IV, and C III in the z = 0.06807 system are consistent with an origin of these species in low density gas photoionized by the extragalactic background. Photoionization modeling implies an ionization parameter logU = -1.35 and a gas-phase abundance of 0.4 (+0.6, -0.2) times solar. The high ionization level of the system requires a low total gas density of $n_H \sim 10^{-5}$

cm$^{-3}$ and an absorption path length of ~80 kpc if the ionization is dominated by photons. If the high states of ionization in this system are instead produced by collisional ionization in hot gas, then the H I must be associated with a cooler phase of a multiphase medium.

   4. The FUSE and new STIS observations confirm the presence of the z = 0.14232 O VI system previously studied by Tripp & Savage (2000) and allow a more accurate assessment of the multi-component nature of this absorber. The system is detected in the lines of Lyα, Lyβ, and O VI λλ1031.93, 1037.62 at z = 0.14232. Upper limits to the absorption line equivalent widths and column density can be placed on various other species at this redshift including C IV, N V, C III, Si III, Si II, and C II. The second strongest Lyα absorption component in this system occurs at z = 0.14262 or 80 km s$^{-1}$ redward of the principal H I component. While this component is detected in C III, it is not evident in the lines of C IV, N V, and O VI.

   5. The multicomponent z = 0.14232 system consists of components containing very highly ionized gas seen only in H I and O VI and moderately ionized components seen in H I and C III. If the O VI absorbing components are produced in gas photoionized by the extragalactic background, the limits on N V and C IV imply logU > -0.74. However, this absorber could also be produced in collisional ionization equilibrium with T > 2.5x10$^5$ K or under conditions of non-equilibrium cooling of hot gas.

   6. The two definite O VI systems toward PG 0953+415 occur at redshifts where the number density of galaxies in the one degree field surrounding the QSO have local maxima. However, the galaxies near the redshifts of the absorbers have projected distances away from the line of sight ranging from 0.4 to 3.5 Mpc.

   7. The FUSE measurements of O VI systems toward PG 0953+415 provide additional evidence for a high number density of O VI absorbers at low redshift. Combining results for PG 0953+415, 3C 273, H 1821+643, and PG 0804+761 we improve the estimate of dN/dz at low redshift. Over a total unobscured redshift path estimated to be Δz ~ 0.43, six O VI systems have been detected with W$_r$ of the λ1031.93 Å line exceeding 50 mÅ implying dN/dz = 14 (+9, -6) for <z> = 0.09. This is a high number density of very highly ionized metal line systems and implies a low redshift value of Ω$_b$(O VI) > 0.002h$_{75}^{-1}$ if the metallicity in the absorbing gas is 0.1 solar or smaller.

    We express our appreciation to the many people who participated in the development and operation of FUSE, STIS, and WIYN. The FUSE observations are based on data obtained for the Guaranteed Time Team by the NASA-CNES-CSA FUSE mission operated by Johns Hopkins University. We thank the STIS team for allowing us to use software they have developed. Special thanks go to William Oegerle for providing many helpful suggestions about how we could improve the manuscript. We thank Buell Jannuzi and Angelle Tanner for the galaxy coordinates required for the WIYN redshift

observations prior to publication. We also appreciate being able to do the photoionization modeling by having access to the program CLOUDY developed by Gary Ferland. This research has used the automated plate scanning catalog of POSS I, which is supported by NSF, NASA, and the University of Minnesota. B.D.S. and T.M.T. acknowledge NASA support through grants GO-8165.02 and GO-8695.01 from the Space Telescope Science Institute. B.D.S. appreciates support from NASA through FUSE contract NAS5-32985. K.R.S. acknowledges partial support from Long Term Space Astrophysics Grant NAG5- 3485.

TABLE 1.
IGM ABSORPTION LINE OBSERVED WITH STIS IN THE z = 0.06807 AND 0.14232 SYSTEMS

| $\lambda_{obs}$ (Å) | z | ID | $\lambda_O$ (Å) | $W_r$ (mÅ) | logN | Mode | Note |
|---|---|---|---|---|---|---|---|
| 1298.42 | 0.06807 | H I Lyα | 1215.67 | 290±13 | >14.06 | E140M | 1 |
| 1323.12 | 0.06805 | N V | 1238.82 | 49±6 | 13.50±0.06 | E140M | 1 |
| 1327.40 | 0.06807 | N V | 1242.80 | 33±6 | 13.54±0.09 | E140M | 1 |
| 1488.63 | 0.06807 | Si IV | 1393.76 | <29 (4σ) | <12.6 | E140M | 2 |
| 1288.63 | 0.06807 | Si III | 1206.50 | <22 (4σ) | <12.1 | E140M | 2 |
| 1346.22 | 0.06807 | Si II | 1260.42 | <40 (4σ) | <12.5 | E140M | 2 |
| 1653.58 | 0.06807 | C IV | 1548.20 | 112±24 | >13.56 | E140M | 1 |
| 1656.31 | 0.06806 | C IV | 1550.77 | 80±18 | >13.86 | E140M | 1 |
| 1425.37 | 0.06807 | C II | 1334.53 | <49 (4σ) | <13.5 | E140M | 2 |
| 1389.00 | 0.14258 | H I Lyα | 1215.67 | 418±28 | 13.99±0.03 | E140M | 3, 4 |
| 1178.78 | 0.14231 | O VI | 1031.93 | 84±15 | 14.04 (+0.10,-0.13) | E140M | 3 |
| 1185.30 | 0.14233 | O VI | 1037.62 | 97±13 | 14.35 (+0.08,-0.10) | E140M | 3 |
| 1415.12 | 0.14233 | N V | 1238.80 | <36 (4σ) | <13.2 | E140M | 3 |
| 1592.13 | 0.14233 | Si IV | 1393.76 | <69 (4σ) | <12.9 | E140M | 3 |
| 1378.22 | 0.14233 | Si III | 1206.50 | <41 (4σ) | <12.3 | E140M | 3 |
| 1490.02 | 0.14233 | Si II | 1304.37 | <27 (4σ) | <13.3 | E140M | 3 |
| 1768.56 | 0.14233 | C IV | 1548.20 | <280 (4σ) | <13.8 | FOS | 5 |
| 1523.47 | 0.14233 | C II | 1334.53 | <42 (4σ) | <13.3 | E140M | 3 |
| 1178.89 | 0.14241 | O VI | 1031.93 | 116±9 | ... | G140M | 6 |
| 1185.45 | 0.14247 | O VI | 1037.62 | 64±11 | ... | G140M | 6 |
| 1171.64 | 0.14226 | HI Lyβ | 1025.72 | 68±15 | ... | G140M | 7 |
| 1768.53 | 0.14232 | C IV | 1548.20 | <64 (4σ) | <13.2 | G230M | 8 |
| 1769.00 | 0.14262 | C IV | 1548.20 | <81 (4σ) | <13.3 | G230M | 8 |
| 1771.48 | 0.14232 | C IV | 1550.77 | <63 (4σ) | <13.5 | G230M | 8 |
| 1771.94 | 0.14262 | C IV | 1550.77 | <70 (4σ) | <13.5 | G230M | 8 |

Notes. (1) These STIS E140M measurements are from Tripp et al. (2002). The values of logN are based on apparent column density integrations. (2) The 4σ upper limits for $W_r$ and logN for the z = 0.06807 system were measured from the STIS E140M spectrum. The limits for logN assume the absorptions fall on a single component COG with b = b(C IV) = 8 km s$^{-1}$. (3) Measurements from Tripp & Savage (2000) based on STIS E140M observations. The listed column densities are based on apparent column density integrations. The 4σ limits assume the absorption is unsaturated and refer to the velocity range from -40 to +40 km s$^{-1}$ in the z= 0.14232 rest frame. (4) We list the wavelength and redshift of the centriod of the H I Lyα absorption which extends over a velocity range of 410 km s$^{-1}$ and contains five definite components. The strongest Lyα component occurs at z = 0.14232. We use that as the reference redshift in velocity plots

appearing in this paper. In that restframe the second strongest H I absorption component occurs at v = 81 km s$^{-1}$.   (5) The large limit for C IV $\lambda$1548.20 from Tripp & Savage (2000) is based on Faint Object Spectrograph Observations (FOS) and is superseded by the higher resolution G230M observations listed at the bottom of the table. (6) Apparent O VI column densities based on the STIS G140M observations have not been determined.  The O VI column density in the z = 0.14132 system based on a COG analysis of the average equivalent widths for all the  O VI absorption line measurements is given in Table 5.  (7) The feature near 1171.6 Å is a blend of weak Ly$\beta$ from the z = 0.14132 system and Ly$\delta$ from the very strong associated Ly$\alpha$ system at z = 0.23349 with logN(HI) = 14.60±0.02 and b = 23±1 based on fits to the Ly$\alpha$ and Ly$\beta$ lines.  (8) For C IV there is no absorption clearly detected at z = 0.14232 or at z = 0.14162.  These equivalent width limits are for integrations  over the velocity ranges -50 to +50 km s$^{-1}$ and from +50  to +150 km s$^{-1}$ in the z = 0.14132 rest frame.  A feature that could possibly be identified as C IV $\lambda$1548.20 at +80 km s$^{-1}$ in the G230M spectrum has W$_r$ = 41±20 mÅ.  The 4$\sigma$ column density limits assume unsaturated absorption.

TABLE 2.  IGM absorption lines observed with FUSE

| $\lambda_{obs}$ (Å) | z | ID | $\lambda_0$ (Å) | LiF1 $W_r$ (mÅ) | LiF2 $W_r$ (mÅ) | Adopted $W_r$ (mÅ) | Note |
|---|---|---|---|---|---|---|---|
| 1038.71 | 0.06804 | H I Lyγ | 972.54  | 60±13  | 66±21  | 60±13  | 1 |
| 1043.52 | 0.06806 | C III   | 977.02  | 91±11  | 70±11  | 91±11  | 2 |
| 1095.53 | 0.06806 | H I Lyβ | 1025.72 | 115±19 | 124±11 | 124±11 | 3 |
| 1102.16 | 0.06806 | O VI    | 1031.93 | 146±15 | 115±12 | 115±12 | 4 |
| 1108.23 | 0.06805 | O VI    | 1037.62 | 93±13  | 81±10  | 81±10  | 5 |
| 1116.39 | 0.14265 | C III   | 977.02  | 31±14  | 22±13  | 26±10  | 6 |
| 1171.50 | 0.14212 | H I Lyβ | 1025.72 | 59±19  | 84±25  | 68±15  | 7 |
| 1178.77 | 0.14230 | O VI    | 1031.93 | 130±18 | 123±23 | 127±14 | 8 |
| 1185.33 | 0.14235 | O VI    | 1037.62 | 91±26  | ...    | 91±26  | 9 |

Notes. (1) The line is identified as H I Lyγ absorption in the system z = 0.06807. The listed rest frame equivalent widths include contamination from an ISM $H_2$ R(2) 5-0 λ1038.689 line estimated to be 26 mÅ in the observed frame based on other $H_2$ J= 2 lines observed in the FUSE spectrum.  We adopt the LiF1 measurement which is superior in resolution and S/N to LiF2.  After correcting for the $H_2$ absorption and transforming to the rest frame we determine $W_r$ (Lyγ) = 36±16 mÅ, where the error includes the systematic uncertainty associated with the deblending of the $H_2$ absorption. (2) The LiF1 measurement is adopted for estimating logN(C III) because there is less fixed-pattern noise than in the LiF2 observation.  (3) We adopt the LiF2  measurement because of  the superior  S/N and resolution and the fact  the LiF1 measurement is near the edge of the detector, which results in an uncertain continuum. (4) We adopt the LiF2 measurement because of its higher resolution and S/N. (5) We adopt the LiF2 measurement because of resolution and S/N.  The values of $W_r$ include a blend from an $H_2$ R(0) 0-0 λ1108.127 line estimated to be 6 mÅ in the observed frame.  The adopted deblended rest frame equivalent width is 76±12 mÅ.  (6) C III λ977.02 is marginally detected near the redshift of the second strongest H I component in the z = 0.14232 system. Combining the two measurements we obtain $W_r$ = 26± 10 mÅ.  No C III is significantly detected at the redshift of the strongest H I component which is near the redshift z = 0.14232 of the O VI absorption. There appears to be a marginal indication of C III at v = 0 km s$^{-1}$ in the LiF1 channel (see Fig. 5).  However, this ~2σ feature is not evident in the LiF2 spectrum.  The 3σ  restframe equivalent width upper limits for C III absorption at z = 0.14232  are 42 and 45 mÅ for LiF1 and LiF2 respectively.  These numbers imply a 4 σ limit of  40 mÅ. (7) The line near 1171.5 Å is a blend of H I Lyδ  at z = 0.23349  and Lyβ in the system

at z = 0.14232. The equivalent width for the blend is listed. The observed frame equivalent width of the blended Lyδ line is estimated to be 45mÅ (see discusion in §4.2). (8) The O VI λ1031.93 absorption line at z = 0.14232 is well detected in the LiF1 and LiF2 channel spectra. We adopt a weighted average of the two observations. A 4σ upper limit for $W_r$ (1031.93) at z = 0.14265 is 72 mÅ. (9) The O VI λ1037.62 absorption line is near the edge of the detector for the LiF1 observation and the continuum placement is uncertain. The O VI λ1037.62 line is off the edge of the detector for the LiF2 channel.

TABLE 3.
O VI, C III, and Lyβ lines at the redshifts of the Lyα Absorbers observed by STIS[a]

| z(Lyα) | $W_r$(Lyα) STIS | Other Lines/Species Definitely Detected[b] | Note[b] |
|---|---|---|---|
| 0.01558 | 58±12 | blend ( OVI) | 1 |
| 0.01602 | 162±13 | ... | |
| 0.01655 | 123±12 | ... | |
| 0.04512 | 37±6 | blend (O VI) | 2 |
| 0.05876 | 280±14 | blend (O VI) | 3 |
| 0.06807 | 290±13 | Lyβ, O VI, N V, C IV, C III | 4 |
| 0.09228 | 68±10 | blend (O VI) | 5 |
| 0.09315 | 174±13 | blend (O VI) | 6 |
| 0.10939 | 168±11 | blend (O VI) | 7 |
| 0.11558 | 115±12 | ... | 8 |
| 0.11826 | 179±10 | ... | 9 |
| 0.11866 | 45±10 | ... | |
| 0.12801 | 94±10 | ... | 10 |
| 0.14258 | 423±23 | Lyβ, O VI, C III | 11 |
| 0.17984 | 84±11 | blend (O VI) | 12 |
| 0.19070 | 74±11 | ... | 13 |
| 0.19142 | 177±15 | ... | |
| 0.19221 | 122±18 | ... | 14 |
| 0.19360 | 269±15 | Lyβ | |
| 0.21514 | 75±9 | blend (O VI) | 15 |
| 0.22527 | 80±9 | blend (O VI) | 16,17 |
| 0.23260 | 172±5 | Lyβ, O VI | 17 |
| 0.23349 | 321±7 | Lyβ, O VI | 17 |

[a] The multi-component Lyα system with lines at z = 0.14178, 0.14233, 0.14263, 0.14294, and 0.14311, is listed as a single system at z = 0.14258 in this table.

[b] The FUSE and STIS spectra were searched for Lyβ, O VI λλ1032.93, 1037.62, C III λ977.02 at the redshifts of the Lyα lines observed by STIS. The 4σ detection limits for the FUSE observations are typically 60mÅ in the rest frame while the STIS 4σ detection limits are typically 40 mÅ. Marginal detections and cases where the possible absorption is strongly blended with other ISM or IGM absorption features are discussed in the notes. ISM and/or IGM blending affecting one or both of the O VI doublet lines is

indicated with the note "blend (O VI)" listed in column 3. The associated systems at $z =$ 0.22527, 0.23260, and 0.23349 are included for completeness.

[b] Notes about absorption lines. (1) $z = 0.01558$. Ly$\beta$ is confused with terrestrial O I emission. O VI $\lambda 1031.93$ blends with ISM Ar I $\lambda 1048.20$. C III $\lambda 977.02$ blends with ISM $H_2$ L9 R1 $\lambda 992.01$. (2) $z = 0.0451$. O VI $\lambda 1037.62$ blends with ISM N II $\lambda 1083.99$. (3) $z = 0.05876$. O VI $\lambda 1031.93$ blends with ISM $H_2$ L1 R1 $\lambda 1092.73$. (4) $z = 0.06807$. The lines detected in this system are discussed in §4.1. (5) $z = 0.09228$. O VI $\lambda 1037.62$ blends with ISM Fe II $\lambda 1133.67$. (6) $z = 0.09315$. A feature of low significance occurs near the expected wavelength of Ly$\beta$. O VI $\lambda 1037.62$ blends with ISM N I $\lambda\lambda 1134.17$, 1134.42. (7) $z = 0.10939$. A feature of low significance occurs near the expected wavelength of Ly$\beta$. The O VI $\lambda 1031.93$ line blends with ISM Fe II $\lambda 1144.94$. C III $\lambda 977.02$ blends with ISM N II $\lambda 1083.99$. (8) $z = 0.11558$. A feature of low significance occurs near the expected wavelength of Ly$\beta$. (9) $z = 0.11826$. A feature of low significance occurs near the expected wavelength of Ly$\beta$. C III $\lambda 977.02$ blends with ISM $H_2$ L1 R1 $\lambda 1092.73$. (10) $z = 0.12801$. C III $\lambda 977.02$ blends with O VI $\lambda 1031.93$ in the $z = 0.06807$ system. (11) $z = 0.14258$. The lines detected in this five or possibly six component Ly$\alpha$ system are discussed in §4.2. (12) $z = 0.17984$. O VI $\lambda 1031.93$ is near the core of strong ISM Ly $\alpha$. O VI $\lambda 1037.62$ blends with Ly$\beta$ at $z = 0.19358$. C III $\lambda 977.02$ blends with ISM PII $\lambda 1152.82$. (13) $z = 0.19070$. O VI $\lambda 1037.62$ blends with Ly $\alpha$ at $z = 0.01602$. (14) $z = 0.19221$. A feature of low significance occurs at the expected wavelength of the O VI $\lambda 1031.92$ line. A corresponding O VI $\lambda 1037.62$ line is not detected. (15) $z = 0.21514$. O VI $\lambda 1031.93$ blends with ISM S II $\lambda 1253.81$. A feature of low significance occurs at the expected wavelength of Ly$\beta$. (16) $z = 0.22527$. O VI $\lambda 1031.93$ blends with Ly$\beta$ at $z = 0.23260$. (17) $z = 0.22527, 0.23260$, and $0.23349$. We consider these systems to be associated systems since their redshifts are near the QSO redshift.

TABLE 4
COLUMN DENSITIES IN THE z = 0.06807 System

| ION | $b^a$ (km s$^{-1}$) | LogN(ion)$^a$ (cm$^{-2}$) | Method$^b$ |
|---|---|---|---|
| H I | 22.5 (+2.8, -2.5) | 14.39 (+0.11, -0.10) | COG for Ly$\alpha$, $\beta$, $\gamma$ |
| H I | 23 ±3 | 14.35 (+0.08, -0.14) | Profile fit to Ly$\alpha$ |
| C II | 8 | < 13.5 (4$\sigma$) | COG with b from C IV |
| C III | 8 (+7, -4) | 13.65 (+1.00, -0.25) | COG with b from C IV |
| C IV | 8 (+7, -4) | 14.03±0.23 | Profile Fit |
| Si II | 8 | < 12.5 (4$\sigma$) | COG with b from C IV |
| Si III | 8 | < 12.1 (4$\sigma$) | COG with b from C IV |
| Si IV | 8 | <12.6 (4$\sigma$) | COG with b from C IV |
| N V | 10±4 | 13.51±0.04 | Profile Fit |
| O VI | 17 (+13, -6) | 14.22 (+0.24, -0.15) | COG |

$^a$ Column density and Doppler parameter errors are 1$\sigma$. Column density limits are 4$\sigma$.
$^b$ The column densites have been derived by Voigt profile fitting and COG techniques. For the STIS observations of H I, C IV, and N V we list the Voigt profile fit results from Tripp et al. (2002). The fitted profiles are shown in Figure 3. For H I we also list the column density and b value based on fitting the STIS Ly$\alpha$ and the FUSE Ly$\beta$, and Ly$\gamma$ equivalent width measurements to a single component Doppler broadened COG. The results for O VI are from a fit of the doublet equivalent widths observed by FUSE to a single component COG. The value of N(C III) assumes b(C III) = b (C IV) = 8(+7, -4) km s$^{-1}$. The 4$\sigma$ upper limits given for C II, Si II, Si III, and Si IV assume b = 8 km s$^{-1}$ which is based on the value found for C IV

TABLE 5. VELOCITIES AND COLUMN DENSITIES IN THE z = 0.14232 SYSTEM

| Species | v (km s$^{-1}$) | b (km s$^{-1}$) | logN (cm$^{-2}$) | Note |
|---|---|---|---|---|
| H I | 0±2 | 31±7 | 13.59±0.03 | 1 |
| O VI | -2±2 | 13(+5, -3) | 14.32(+0.19,-0.13) | 2 |
| N V | -40 to 40 | ... | <13.2 (4_) | 3 |
| Si IV | -40 to 40 | ... | <12.9 (4_) | 3 |
| Si III | -40 to 40 | ... | <12.3 (4_) | 3 |
| C IV | -50 to 50 | ... | <13.2 (4_) | 4 |
| C III | -50 to 50 | ... | <12.8 (4_) | 4 |
| H I | 81±2 | 30±9 | 13.43±0.04 | 1 |
| C III | ~ 80 | ... | 12.6± 0.2: | 5 |
| C IV | 50 to 150 | ... | <13.3 (4_) | 4 |
| O VI | 50 to 150 | ... | <13.8 (4_) | 4 |
| H I | -142±3 | 12(+15, -7) | 12.74±0.10 | 1 |
| H I | 163±7 | ~20 | 12.6±0.21 | 1 |
| H I | 207±3 | 19 (+15, -9) | 13.08±0.08 | 1 |
| H I | 269±4 | 17 (+26, -10) | 12.82±0.12 | 1 |

Notes. (1) Values of v, b, and logN for H I are derived from the Voigt profile fits to the Lyα line given by Tripp & Savage (2000). The six H I components span a total velocity range of 410 km s$^{-1}$. Metal lines are detected in the two strongest H I components at 1 and 81 km s$^{-1}$. The weak component at 163 km s$^{-1}$ is only marginally detected. (2) The value of logN and b for O VI were derived by fitting a single component COG to the values of W$_r$(1031.93) = 112±7 mÅ and W$_r$(1037.62) = 81±8 mÅ obtained from an error weighted average of the STIS E140M, STIS G140M, and FUSE observations. The velocity is based on the simultaneous Voigt profile fit to the STIS E140M observations of the O VI λλ1031.93, 1037.62 doublet reported by Tripp & Savage (2000). (4) These 4σ limits assume unsaturated absorption. (5) The value of logN(C III) reported here was obtained from a weighted average of the LiF1 and LiF2 observations listed in Table 2 , W$_r$(977.02) = 26±10 mÅ, and the assumption the absorption line is on the linear part of the COG. The result is very uncertain since it is based on a 2.6σ detection.
27

TABLE 6. GALAXIES WITHIN ~200 km s$^{-1}$ OF THE z = 0.06807 ABSORBER

| Galaxy Redshift[a] | $\rho$[b] (Mpc) | $\Delta v$[c] (km s$^{-1}$) | R.A.[d] (J2000) | Decl.[d] (J2000) | $O_{APS}$[e] | $(O-E)_{APS}$[e] | $B_J$[f] | $M_B$[g] |
|---|---|---|---|---|---|---|---|---|
| 0.06792 | 3.48 | -42 | 9 59 47.83 | 40 39 45.6 | 18.3 | 1.7 | 17.8 | -19.6 |
| 0.06803 | 2.00 | -11 | 9 58 19.01 | 40 52 49.3 | 16.8 | 1.7 | 16.2 | -21.2 |
| 0.06861 | 2.12 | 152 | 9 54 25.04 | 41 05 48.7 | 18.0 | 1.6 | 17.5 | -19.9 |
| 0.06867 | 1.09 | 168 | 9 55 34.18 | 41 11 49.1 | 17.9 | 1.3 | 17.4 | -20.0 |
| 0.06882 | 0.917 | 211 | 9 56 36.33 | 41 03 04.3 | 18.8 | 1.6 | 18.3 | -19.1 |

[a] Internal redshift errors are estimated by Tripp et al. (1998) to be ~ 50 km s$^{-1}$.

[b] Projected distance to the sight line (impact parameter) assuming $H_o$ = 75 km s$^{-1}$ Mpc$^{-1}$ and $q_o$ = 0.0. The coordinates (J2000) of PG 0953+415 are R.A. = 9$^h$ 56$^m$ 52.4$^s$, Dec. = +41° 15' 22.0".

[c] $\Delta v = c (z_{gal} - z_{abs}) / (1 + z_{mean})$ where $z_{mean}$ is the mean of $z_{abs}$ and $z_{gal}$.

[d] Units of right ascension are hours, minutes, and seconds, and the units of declination are degrees, arcminutes, and arcseconds.

[e] POSS I O and E magnitudes from the revised automated plate scanner catalog (Pennington et al. 1993).

[f] $B_J$ magnitude calculated from $O_{APS}$ -18 = 0.96 ($B_J$ -18) +0.52 from Odewahn & Aldering (1995).

[g] Absolute magnitude calculated using the extinction correction, E(B-V) = 0.02, from Lockman & Savage (1995) and the K-correction K = 2.5log(1+z).

TABLE 7
SUMMARY OF LOW z O VI SYSTEM DETECTIONS WITH $W_r(1031.93) > 50$ mÅ

| QSO | z(QSO) | $\Delta z^a$ | $n(Ly\alpha)^b$ | $n(O\ VI)^c$ | $\Sigma N(O\ VI)^d$ | References[e] |
|---|---|---|---|---|---|---|
| H1821+643 | 0.297 | 0.17 | 25 | 4 | $4.2 \times 10^{14}$ | 1, 2 |
| PG 0953+415 | 0.239 | 0.13 | 20 | 2 | $4.3 \times 10^{14}$ | 3, 4 |
| PG 0804+761 | 0.102 | 0.050 | $\geq 1^f$ | 0 | 0 | 5 |
| 3C 273 | 0.158 | 0.083 | 12 | 0 | 0 | 6, 7 |
| PKS 0405-1219 | 0.573 | ...[g] | 19 | $\geq 1^g$ | $\geq 5.8 \times 10^{14}$ | 8, 9 |

[a] Total absorption path approximately corrected for spectral regions blocked by ISM absorption or absorption from unrelated IGM features. An inspection of Table 3 reveals the seriousness of the blockage problem. Of the 20 Lyα absorbing systems detected between redshift 0.01558 and 0.21514 in the spectrum of PG 0953+415, one or both of the O VI lines blends with ISM or IGM absorption systems from other redshifts in 8 cases. In some cases the blending problem severe, while in other cases the blending is modest and we would be able to see a strong O VI IGM feature if present. Accurately allowing for this blockage problem is beyond the scope of this paper. Here we approximately account for the blockage by assuming that the results for PG 0953+415 are representative and that blockage affects 8/20 or ~40% of the systems and ~40% of the redshift path to each QSO. The values of $\Delta z$ listed are therefore taken to be $\Delta z = 0.60\ [0.983\ z(QSO)-0.017]$, where the redshift path is assumed to extend to 5000 km s$^{-1}$ of the QSO. In the case of PG 0804+761, this approximate method of calculating the blockage gives a result ($\Delta z = 0.050$) consistent with the more accurate estimate ($\Delta z = 0.054$) of Richter et al. (2001a).

[b] Number of intervening Lyα systems detected with $W_r > 50$ mÅ. For PG 0953+415 the multi-component Lyα system centered near $z = 0.14258$ (see Table 3) is counted as a single system. In the case of PKS 0405-1219, the Lyα lines are from HST Faint Object Spectrograph observations (Januzzi et al. 1998) and have limiting equivalent widths of ~250 mÅ.

[c] Number of intervening O VI absorption line systems seen in STIS and FUSE observations with $W_r$ of the $\lambda$1031.93 line exceeding 50 mÅ. Additional weaker O VI systems with $W_r < 50$ mÅ have been detected in the STIS spectrum of H1821+643 (2 systems, Tripp et al. 2000) and the FUSE spectrum of 3C 273 (2 systems, Sembach et al. 2001b).

[d] Total O VI column density summed over the measured O VI column densities found in the definite absorbers with $W_r$ (1031.93) > 50 mÅ.

[e] References: (1) Tripp et al. 2000; (2) Oegerle et al. 2000; (3) Tripp & Savage 2000; (4) this paper; (5) Richter et al. 2001a, (6) Sembach et al. 2001b; (7) Morris et al. 1991; (8) Chen & Prochaska (2001); Jannuzi et al. (1998).

[f] GHRS observations of PG0804+761 only cover a small redshift window for H I Ly$\alpha$ absorption. Additional Ly$\alpha$ lines may exist.

[g] Chen & Prochaska (2001) give results based on the analysis of one high column density O VI system toward PKS 0405 -1219 at z = 0.167. However, they did not report on a search for other possible O VI systems over the complete redshift path covered by the STIS spectrum.

FIGURES

FIG. 1- FUSE spectra for PG 0953+415 covering the wavelength range from 915 to 1187 Å. For $\lambda < 1000$ Å the data from the two SiC channels are plotted. For $\lambda > 1000$ Å the data for the LiF channels are plotted except for the region from 1075 to 1093 where the SiC measurements are shown. The spectra are binned by six FUSE pixels corresponding to ~ 0.04 Å for display purposes. The ISM and IGM line identifications are listed to the right of each panel. When a question mark is added to the listed wavelength the feature does not appear in the spectra from two or more channels at a significance level exceeding $3\sigma$. ISM molecular hydrogen lines are identified by a notation system where W1R2 986.241 represents the Werner (1-0) R2 $\lambda$986.241 line and L9P1 992.808 represents the Lyman (9-0) P1 $\lambda$992.808 line. The molecular hydrogen wavelengths are from Abgrall et al. (1993a,b). In order to correct reduced count rates due to object misalignments in the apertures, the SiC1B flux values were scaled to match the flux recorded in the other channels.

FIG. 2- Continuum normalized absorption line profiles are shown on a restframe velocity basis for the FUSE observations of H I Ly$\beta$, Ly$\gamma$, C III $\lambda$977.02, and O VI $\lambda\lambda$1031.93, 1037.62 in the $z = 0.06807$ system. Measurements from two channels are displayed. Depending on the wavelength, the signal-to-noise and resolution are generally superior in one of the two channels displayed. Equivalent widths are listed in Table 2. The observations are binned to ~7 km s$^{-1}$ wide pixels.

FIG. 3-. Continuum normalized absorption line profiles are shown on a restframe velocity basis for the STIS E140M observations of H I Ly$\alpha$, N V $\lambda\lambda$1238.82, 1242.80, C IV $\lambda\lambda$1548.20, 1550.77, Si III $\lambda$1206.50, C II $\lambda$1334.53, and Si II $\lambda$1260.42 in the $z = 0.06807$ system. Voigt profile fits are shown for the detected species. The resolution of these observations is ~ 7 km s$^{-1}$. Equivalent widths are listed in Table 1. The spectra are binned to ~ 3.3 km s$^{-1}$ wide pixels.

FIG. 4- A photoionization model for the O VI system at $z = 0.06807$ is displayed. LogN(ion) is plotted against log U, the ionization parameter, for a constant-density slab illuminated by the extragalactic UV background radiation field of Haardt & Madau (1996) appropriate for $z = 0.06$. The top axis shows the value of the total hydrogen density, log $n_H$ (cm$^{-3}$). The metallicity assumed in the model is [M/H] = -0.4 and the relative abundances among the heavy elements are assumed to be in the Solar ratio. The model calculation assumes logN(H I) = 14.39. The solid lines show the model predictions for the ions indicated. The data points are the column densities of the species listed in Table 4. The observations are consistent with a value of logU = -1.35 and an origin along a 80 kpc path through low density photoionized gas with a metallicity of [M/H] = -0.4 dex and $n_H$ ~ 10$^{-5}$ cm$^{-3}$.

FIG. 5- Continuum normalized absorption line profiles are shown on a restframe velocity basis for the FUSE observations of H I Ly$\beta$, Ly$\gamma$ (no detection), C III $\lambda$977.02, and O VI $\lambda\lambda$1031.93, 1037.62 in the z = 0.14232 system. The observations have been binned to ~ 7 km s$^{-1}$ wide pixels.

FIG. 6. Continuum normalized absorption line profiles are shown on a restframe velocity basis for the STIS G140M observations of H I Ly$\beta$, O VI $\lambda\lambda$1031.93, 1037.62 in the z = 0.14232 system. The STIS G230M measurements in the region of C IV $\lambda\lambda$1548.22, 1550.77 are also shown. The presence of O VI but absence of C IV near v = 0 km s$^{-1}$ indicates a high state of ionization in the O VI system. At the top of the figure we also display the STIS E140M observation of Ly$\alpha$ from Tripp & Savage (2000). The G140 and G230M observations have a resolution of 30 km s$^{-1}$ while the E140M observation has a resolution of 7 km s$^{-1}$. The G140M and G230M observations have been binned to 12-13 km s$^{-1}$ wide pixels.

FIG. 7- A photoionization model for the O VI system at z = 0.14232 is displayed. LogN(ion) is plotted against log U for a constant-density slab illuminated by the extragalactic UV background radiation field of Haardt & Madau (1996) appropriate for z = 0.12. The top axis shows the value of the total hydrogen density, log $n_H$ (cm$^{-3}$). The metallicity assumed in the model is [M/H] = -0.4 and the relative abundances among the heavy elements are assumed to be in the Solar ratio. The model calculation assumes logN(H I) = 13.59. The solid lines show the model predictions for the ions indicated. The data points are the column densities or limits of the species listed in Table 5 for the absorption in the H I and O VI components near v = 0 km s$^{-1}$ in the z = 0.14232 rest frame. The observations are consistent with a value of logU > -0.74 and an origin along a > 420 kpc path through low density photoionized gas. For logU = -0.74, the required metallicity and density are [M/H] ~ -0.4 dex and $n_H$ ~ 10$^{-5.7}$ cm$^{-3}$. For logU = -0.1, these numbers change to [M/H] ~ -0.65 dex and $n_H$ ~ 10$^{-6.3}$ cm$^{-3}$.

FIG. 8- Number of galaxies found in the one degree diameter field centered on PG 0953+415 versus redshift. The galaxy redshifts were obtained by observations with the multiple object spectrograph at the WIYN observatory. The vertical lines above the histogram denote the redshifts of Ly$\alpha$ absorbers (lower set of lines) and the O VI absorbers (two upper lines), with the height of the lines proportional to the absorption column densities as indicated in the Key. The two O VI systems occur at redshifts where there are peaks in the redshift distribution of galaxies in the field of PG 0953+415. O VI absorption is not detected at the redshift of the very strong peak in the galaxy distribution at z ~ 0.042.

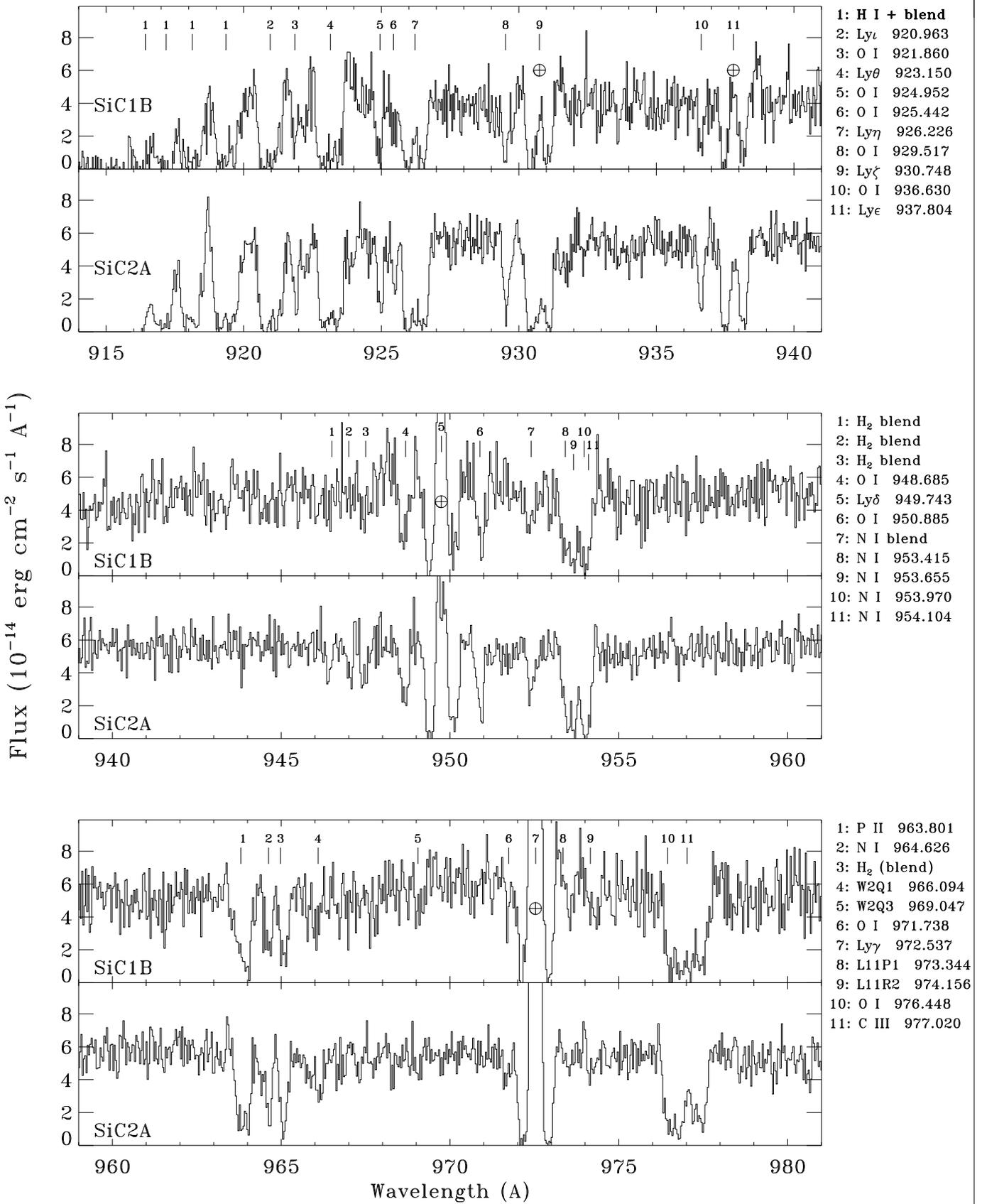

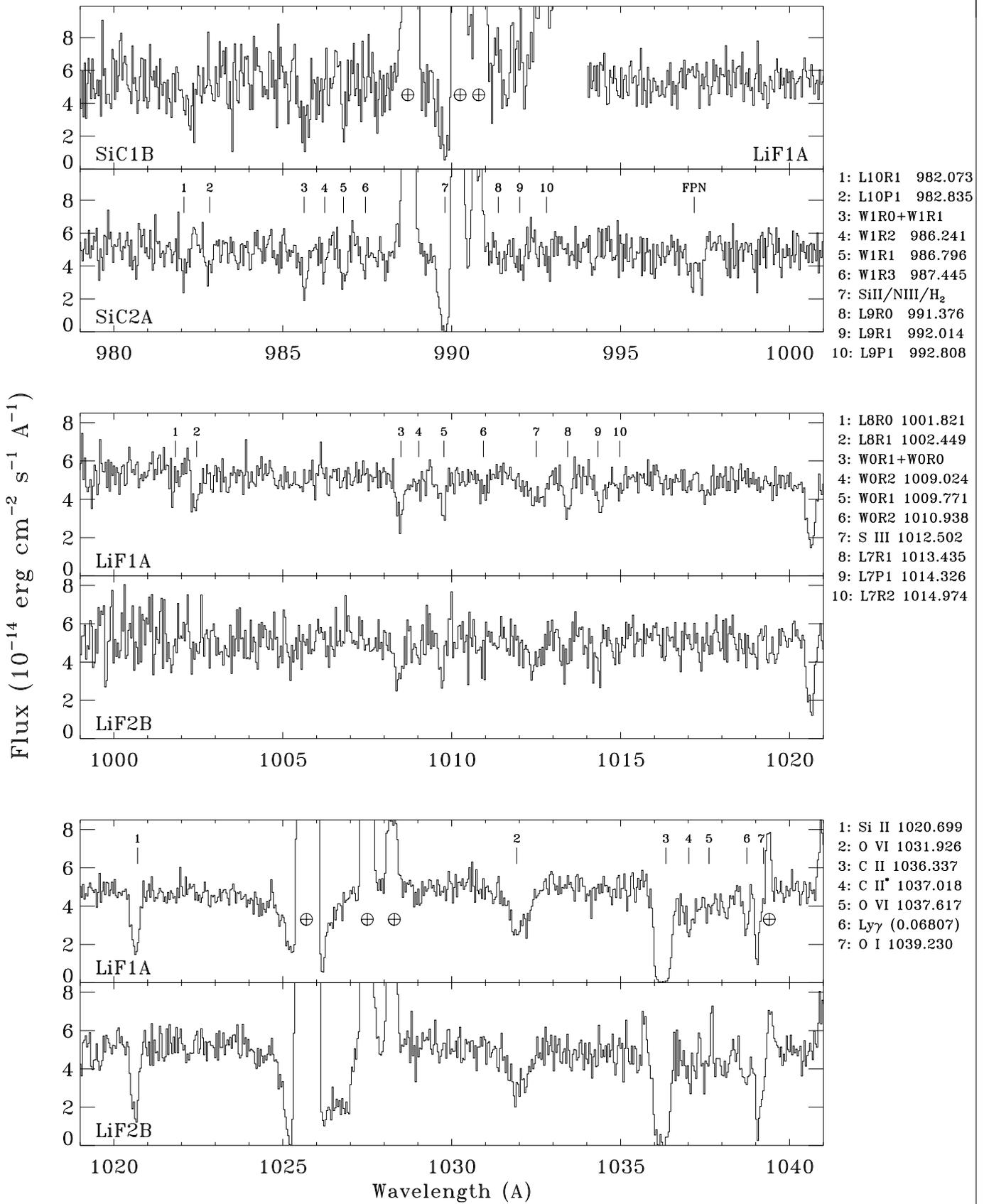

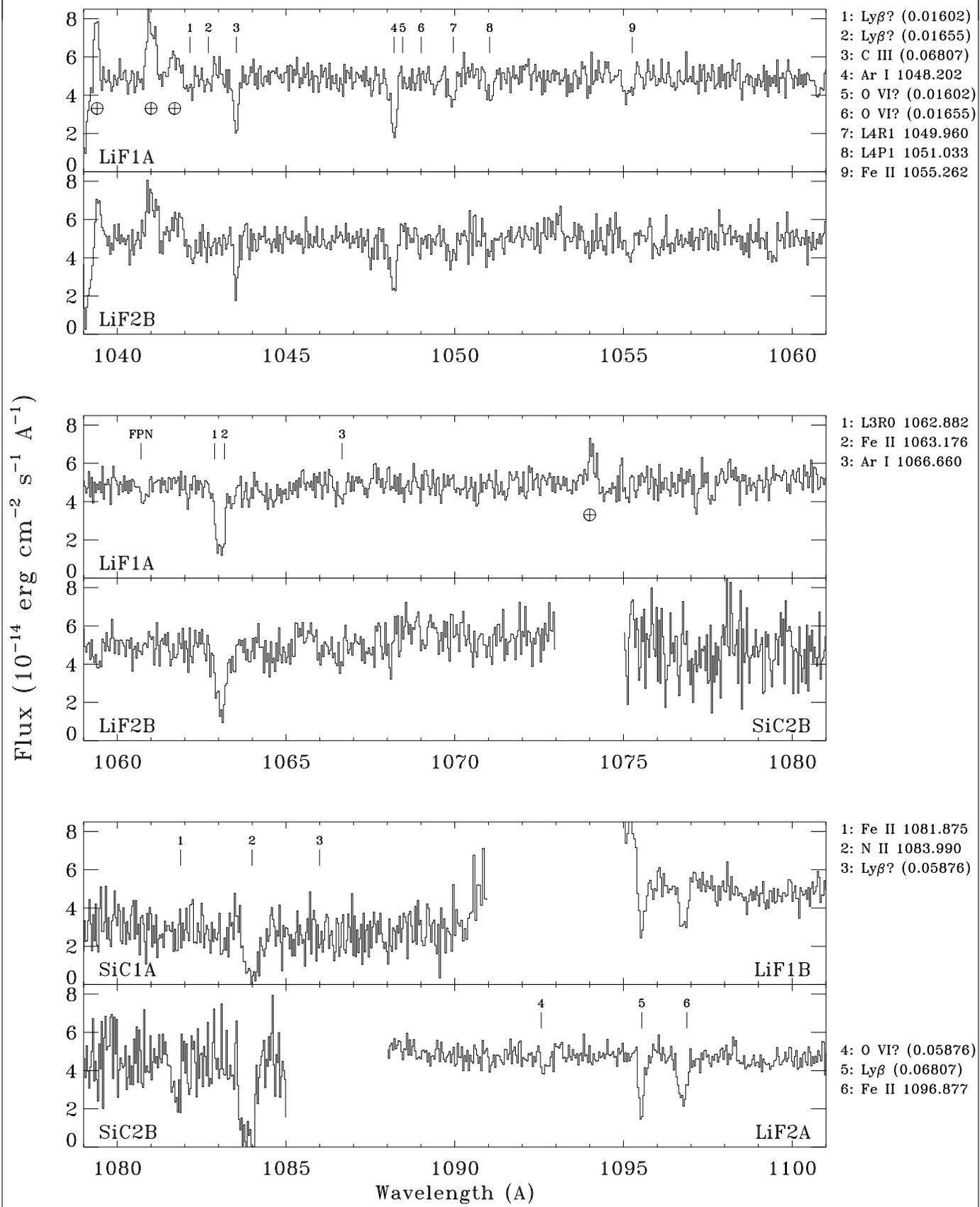

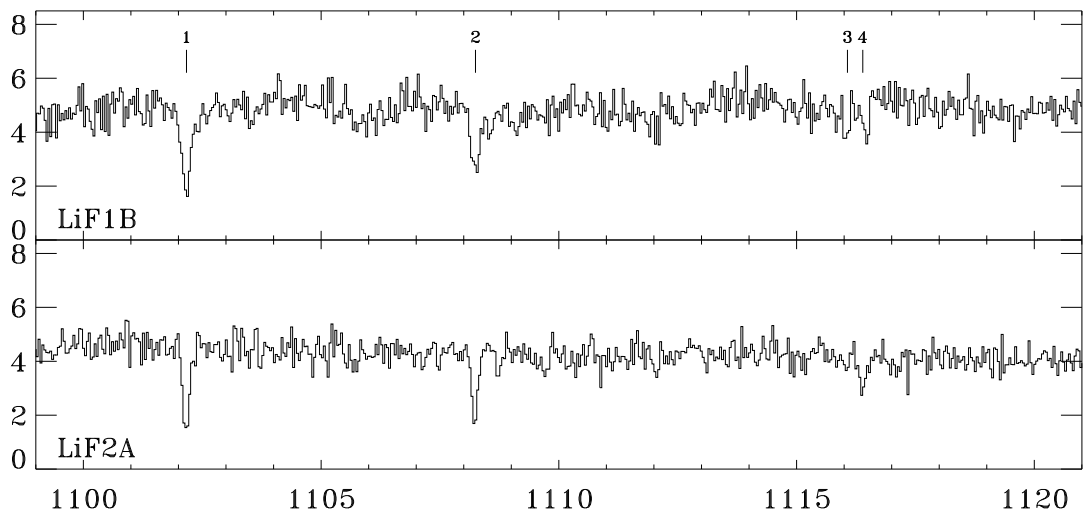
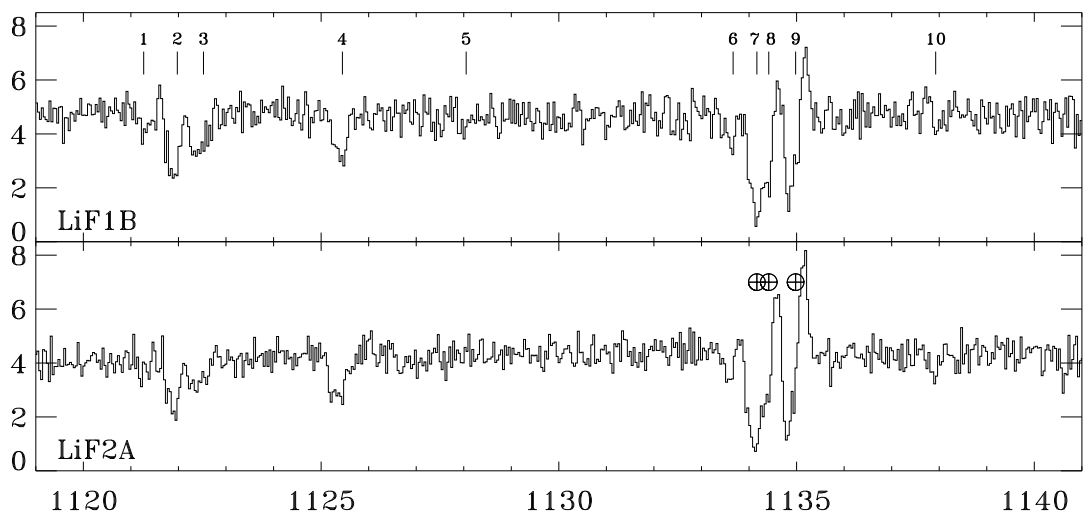
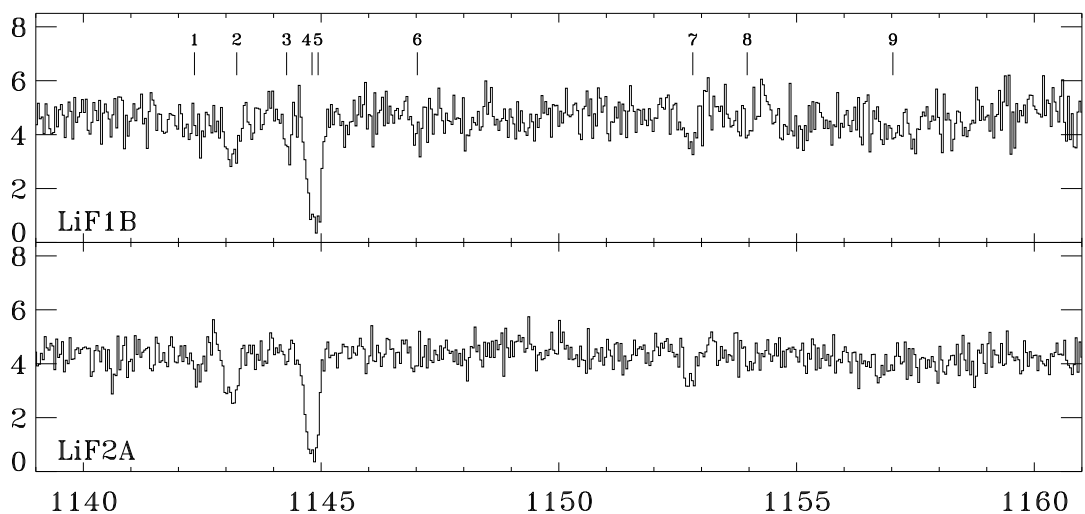

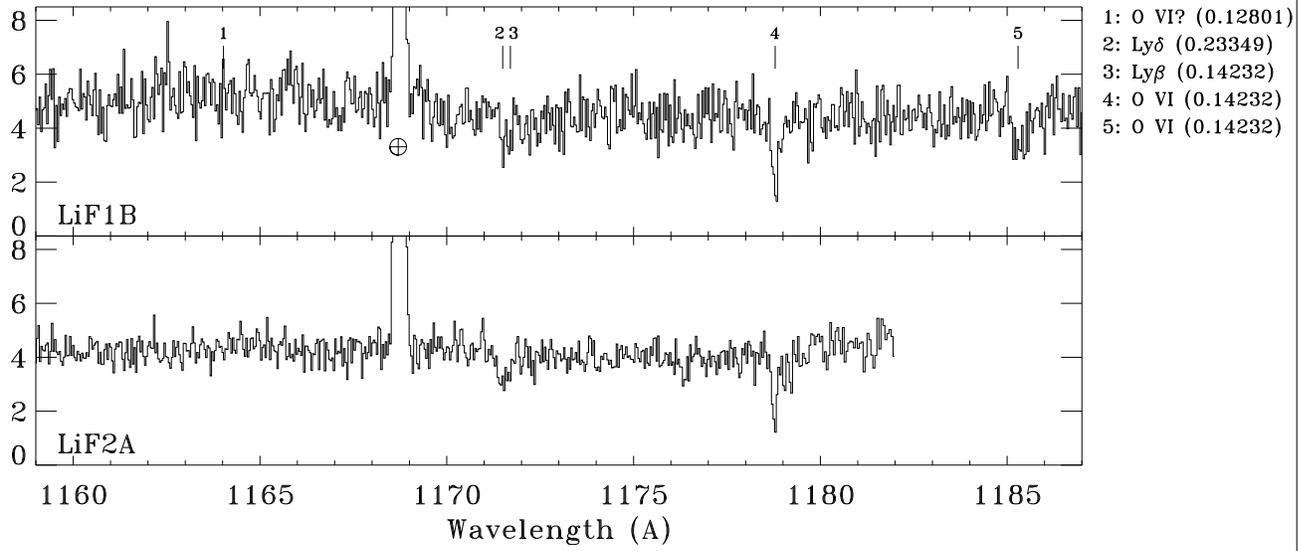

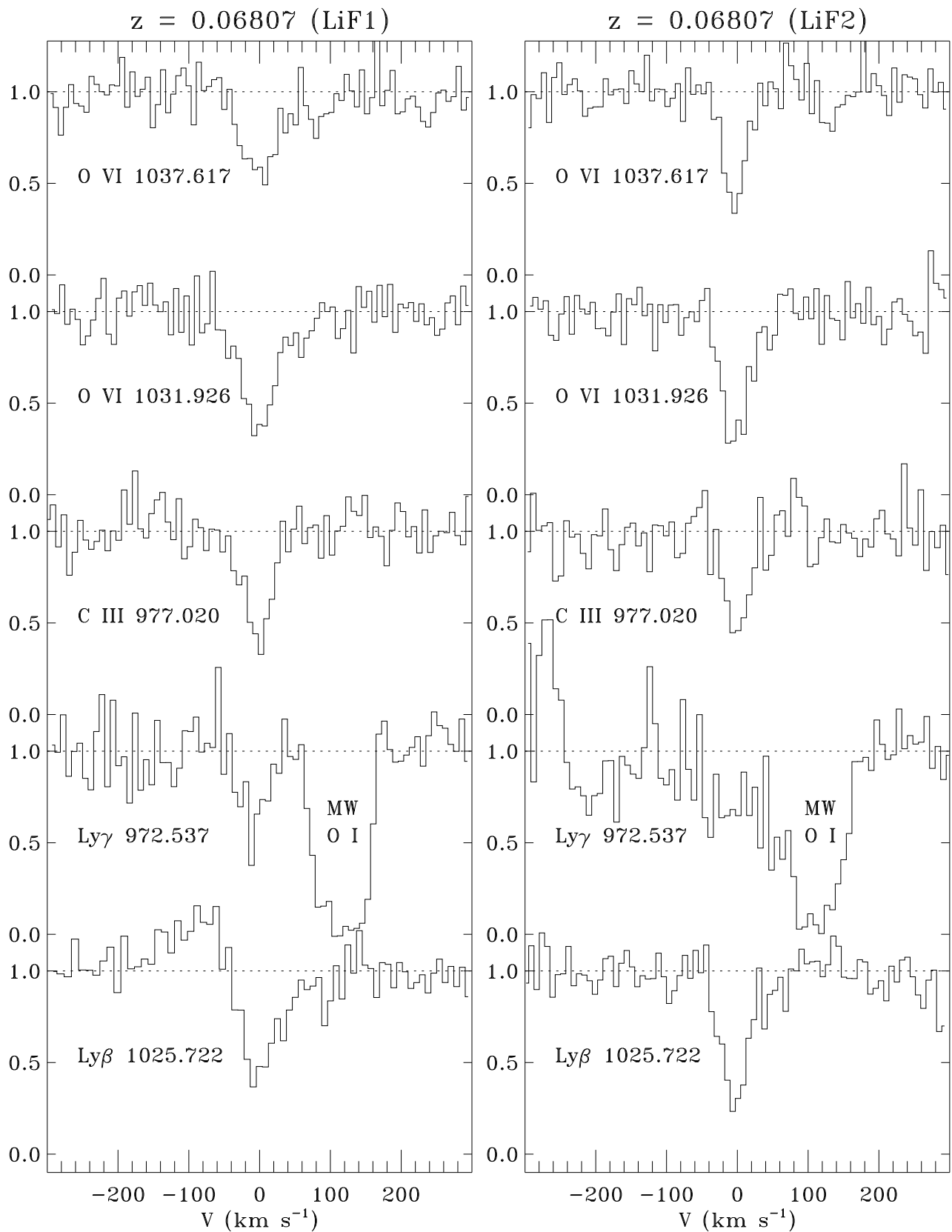

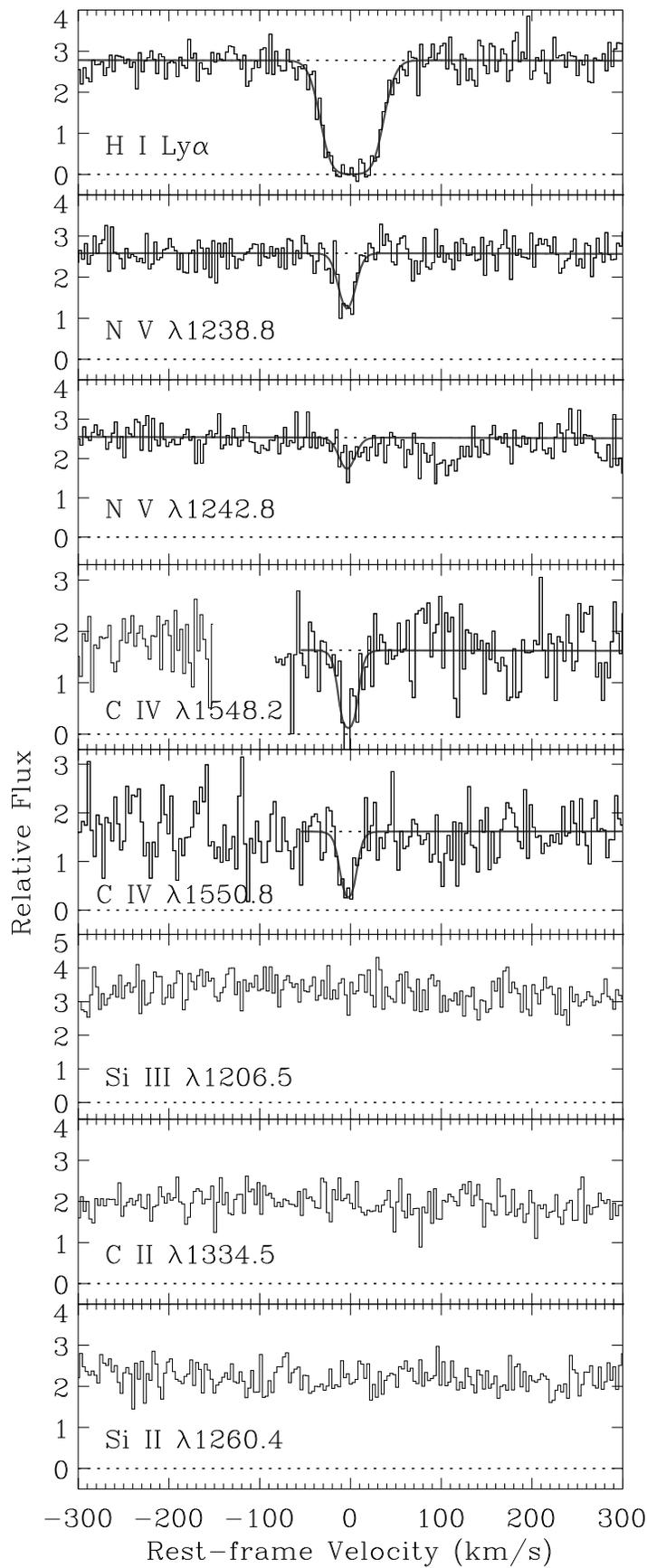

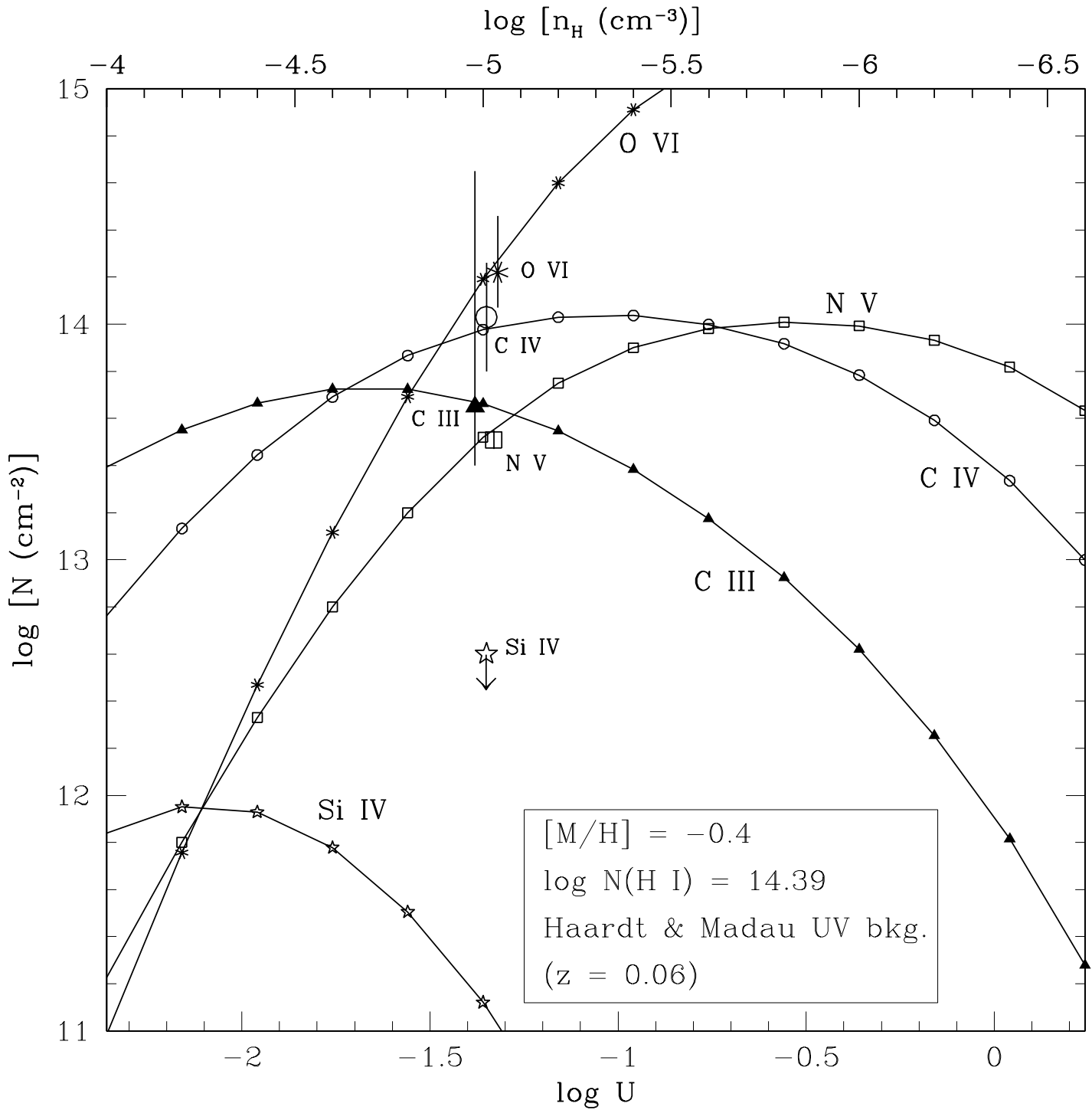

# PG0953+415 (FUSE)

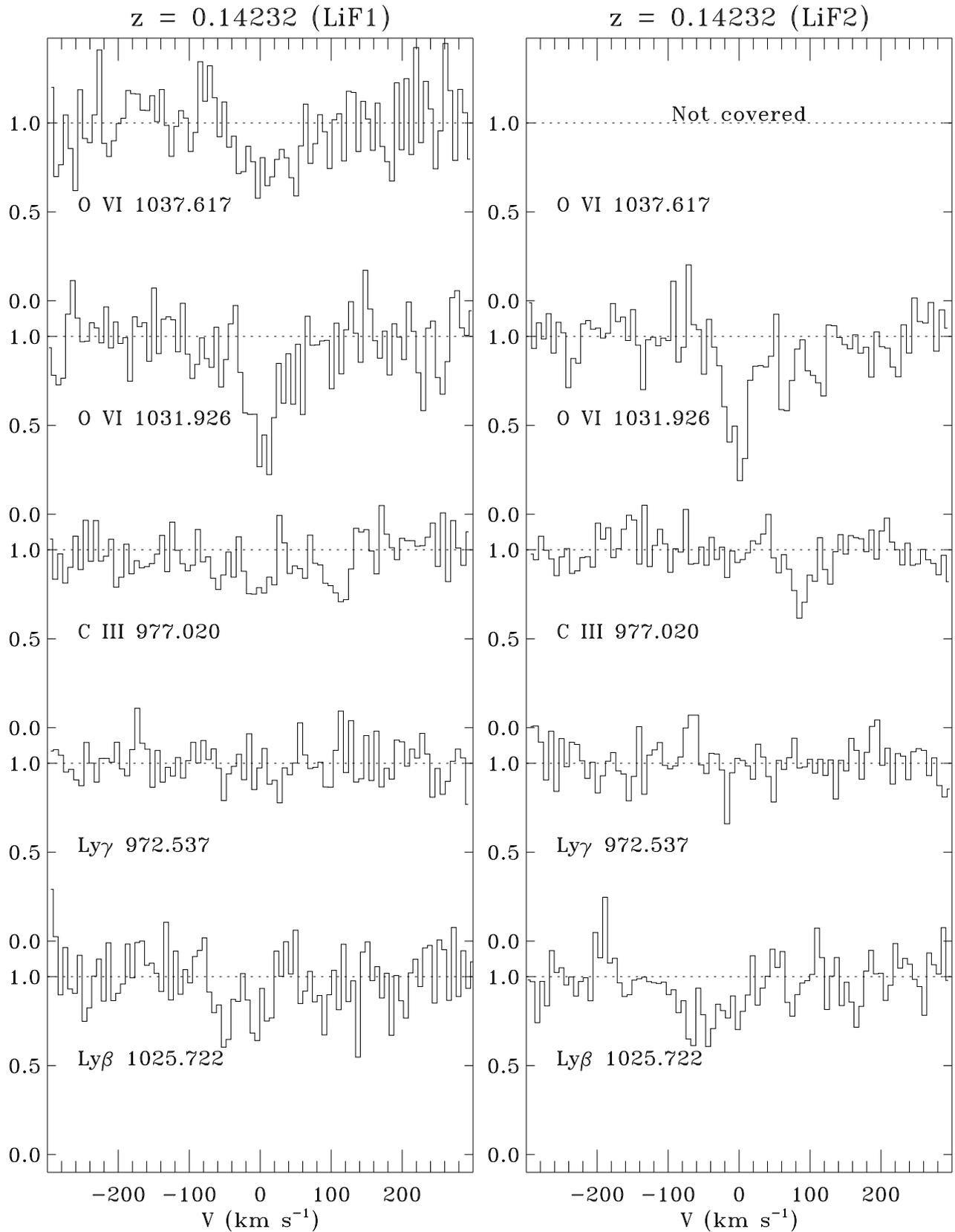

z=0.14232

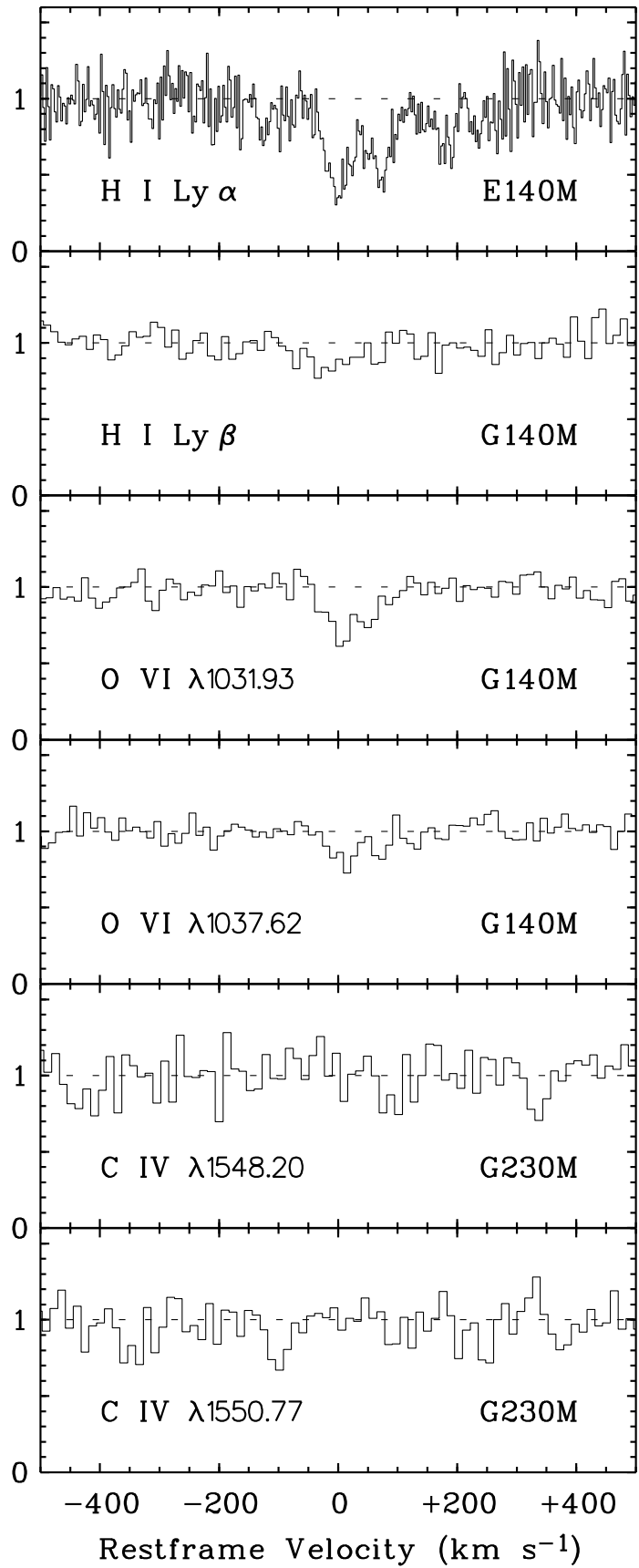

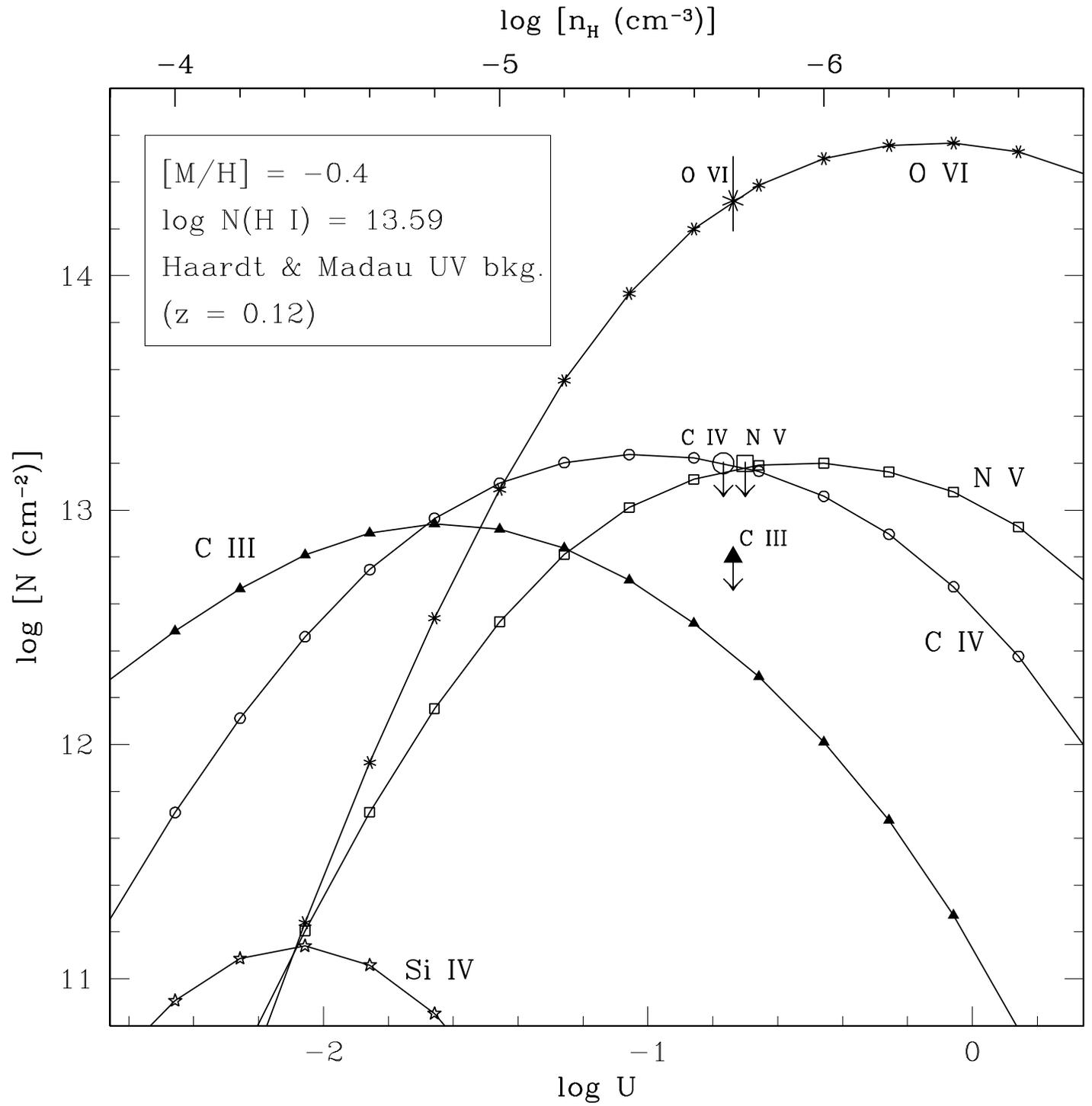

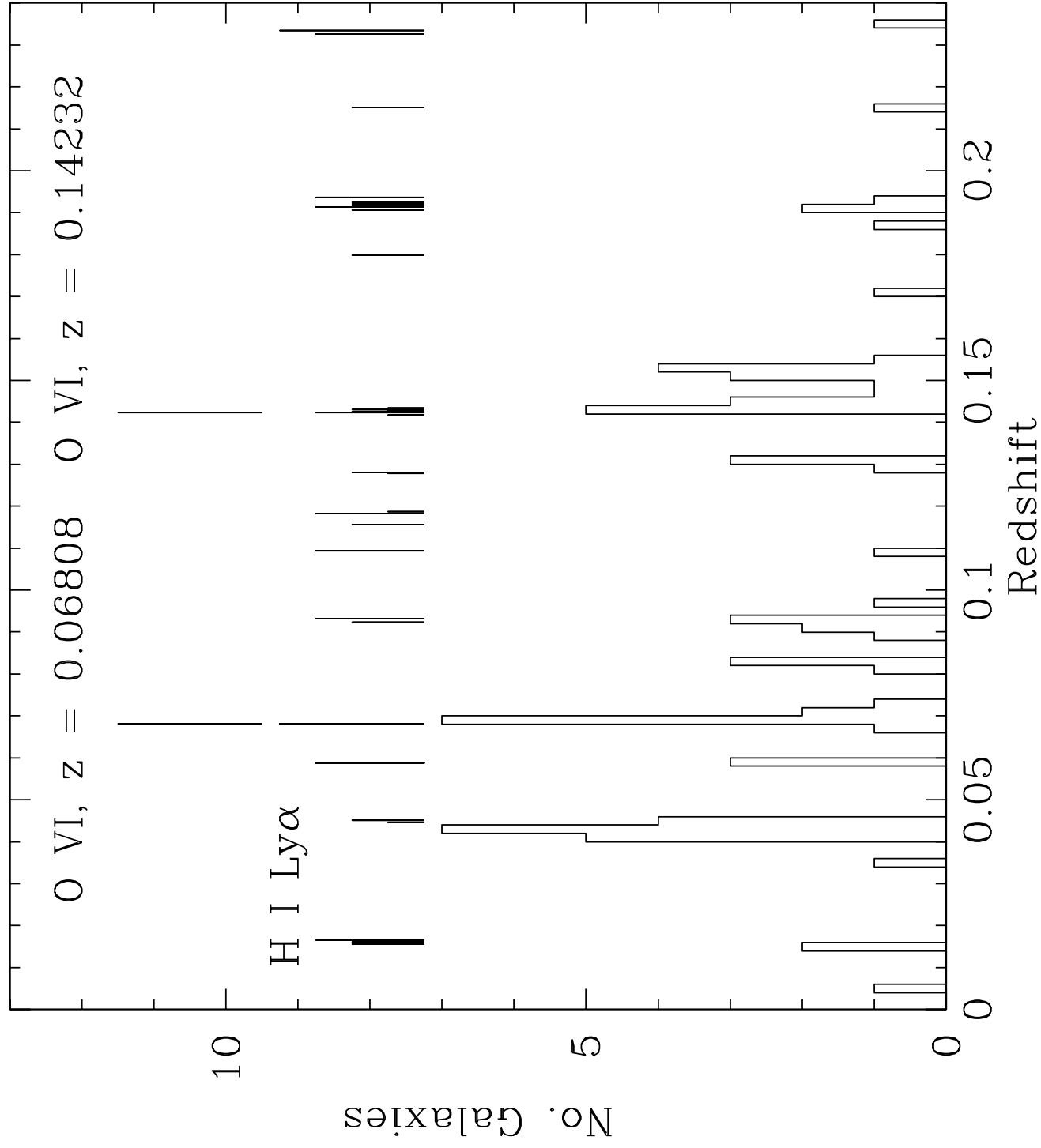